\documentclass[universe,article,accept,pdftex,moreauthors]{Definitions/mdpi} 

\firstpage{1} 
\pubvolume{1}
\issuenum{1}
\articlenumber{0}
\pubyear{2024}
\copyrightyear{2024}
\externaleditor{Academic Editor: }
\datereceived{18 July 2024} 
\daterevised{16 August 2024} % Comment out if no revised date
\dateaccepted{20 August 2024} 
\datepublished{ } 
\hreflink{https://doi.org/} % If needed use \linebreak
\usepackage{amsmath}
\usepackage{amssymb}
\usepackage{threeparttable}
\usepackage{subcaption}
\usepackage{graphicx}

\makeatletter
\newcommand{\Rmnum}[1]{\expandafter\@slowromancap\romannumeral #1@}
\makeatother

\Title{Minute-Cadence Observations of the LAMOST Fields with the TMTS: IV--- %MDPI: Periods (full-stops) are not allowed in the middle or at the end of the article title. A colon or em dash is recommended instead of a period.
 Catalog of Cataclysmic Variables from the \mbox{First 3-yr %MDPI: please confirm if the unit could be y or years. if yes, please check and revise the whole paper.
 Survey}} %Attention: Title altered. Please check that the intended meaning has been retained.

\TitleCitation{Minute-Cadence Observations of the LAMOST Fields with the TMTS: IV---Catalog of Cataclysmic Variables from the First 3-yr Survey}

\Author{Qichun %MDPI: Please carefully check the accuracy of names and affiliations.
 Liu $^{1}$,  Jie Lin $^{1,2,3,}$*, Xiaofeng Wang $^{1,}$*, Zhibin Dai $^{4,5}$, Yongkang Sun $^{6,7}$, Gaobo Xi $^{1}$, Jun Mo $^{1}$, Jialian Liu $^{1}$, Shengyu Yan $^{1}$, Alexei V. Filippenko $^{8}$, Thomas G. Brink $^{8}$, Yi Yang $^{1,8}$, Kishore C. Patra $^{8}$, Yongzhi Cai $^{4,5,9}$,\linebreak Zhihao Chen $^{1}$, Liyang Chen $^{1}$, Fangzhou Guo $^{1}$, Xiaojun Jiang $^{7,10,11}$, Gaici Li $^{1}$, Wenxiong Li $^{10}$, Weili Lin $^{12}$,\linebreak Cheng Miao $^{1}$, Xiaoran Ma $^{1}$, Haowei Peng $^{1}$, Qiqi Xia $^{1}$, Danfeng Xiang $^{1}$ and Jicheng Zhang $^{13}$}

\AuthorNames{Qichun Liu,  Jie Lin, Xiaofeng Wang, Zhibin Dai, Yongkang Sun, Gaobo Xi, Jun Mo, Jialian Liu, Shengyu Yan, Alexei V. Filippenko, Thomas G. Brink, Yi Yang, Kishore C. Patra, Yongzhi Cai, Zhihao Chen, Liyang Chen, Fangzhou Guo, Xiaojun Jiang, Gaici Li, Wenxiong Li, Weili Lin, Cheng Miao, Xiaoran Ma, Haowei Peng, Qiqi Xia, Danfeng Xiang and Jicheng Zhang }

\AuthorCitation{Liu, Q.; Lin, J.; Wang, X.; Dai, Z.; Sun, Y.; Xi, G.; Mo, J.; Liu, J.; Yan, S.; Filippenko, A.V.;~et~al.}
\address{%
$^{1}$ \quad Physics Department and Tsinghua Center for Astrophysics, Tsinghua University, Beijing 100084, China; lqc22@mails.tsinghua.edu.cn (Q.L.); hgb18@mails.tsinghua.edu.cn (G.X.); moj20@mails.tsinghua.edu.cn (J.M.); liu-jl22@mails.tsinghua.edu.cn (J.L.); yansy19@mails.tsinghua.edu.cn (S.Y.); yi.yang@berkeley.edu (Y.Y.); chenzh18@mails.tsinghua.edu.cn (Z.C.); chenly23@mails.tsinghua.edu.cn (L.C.); \mbox{gfz20@mails.tsinghua.edu.cn (F.G.);} lgc21@mails.tsinghua.edu.cn (G.L.); \mbox{miaoc19@mails.tsinghua.edu.cn (C.M.);} maxr20@mails.tsinghua.edu.cn (X.M.); phw23@mails.tsinghua.edu.cn (H.P.); xiaqiqi@mail.tsinghua.edu.cn (Q.X.); \mbox{xiangdf@mail.tsinghua.edu.cn (D.X.)} %MDPI: We added the email addresses here according to those submitted online at susy.mdpi.com. Please confirm.
\\
$^{2}$ \quad CAS Key Laboratory for Research in Galaxies and Cosmology, Department of Astronomy, University of Science and Technology of China, Hefei 230026, China\\
$^{3}$ \quad School of Astronomy and Space Sciences, University of Science and Technology of China, Hefei 230026, China\\
$^{4}$ \quad Yunnan Observatories, Chinese Academy of Sciences, Kunming 650216, China; zhibin\_dai@ynao.ac.cn (Z.D.); caiyongzhi@ynao.ac.cn (Y.C.) %MDPI: 
\\
$^{5}$ \quad Key Laboratory for the Structure and Evolution of Celestial Objects, Chinese Academy of Sciences,\linebreak Kunming 650216, China\\
$^{6}$ \quad National %MDPI: please confirm if the address information in aff 6 and 10 is same, if yes, please merge them.
 Astronomical Observatories, Chinese Academy of Sciences, Beijing 100101, China; sunyk@bao.ac.cn %MDPI: 
\\
$^{7}$ \quad School of Astronomy and Space Science, University of Chinese Academy of Sciences, Beijing 100049, China; xjjiang@bao.ac.cn %MDPI: 
\\
$^{8}$ \quad Department of Astronomy, University of California, Berkeley, CA 94720-3411, USA; \mbox{afilippenko@berkeley.edu (A.V.F.); tgbrink@berkeley.edu (T.G.B.); kcpatra@berkeley.edu (K.C.P.)} %MDPI: 
\\
$^{9}$ \quad International Centre of Supernovae, Yunnan Key Laboratory, Kunming 650216, China\\
$^{10}$\quad CAS Key Laboratory of Optical Astronomy, National Astronomical Observatories, Chinese Academy of Sciences, Beijing 100101, China; li-wx15@tsinghua.org.cn %MDPI: 
\\
$^{11}$\quad Center for Astronomical Mega-Science, Chinese Academy of Sciences, 20A Datun Road, Chaoyang District, Beijing 100101, China\\
$^{12}$\quad Department of Astronomy, Xiamen University, Xiamen 361005, China; linwl@xmu.edu.cn %MDPI: 
\\
$^{13}$\quad Department of Astronomy, Beijing Normal University, Beijing 100875, China; jczhang@bnu.edu.cn %MDPI: 
\\
}

\corres{Correspondence: 
linjie2019@ustc.edu.cn (J.L.); 
wang\_xf@mail.tsinghua.edu.cn (X.W.). }

\abstract{The Tsinghua University--Ma Huateng Telescopes for Survey (TMTS) started to monitor the LAMOST plates in 2020, leading to the discovery of numerous short-period eclipsing binaries, peculiar pulsators, flare stars, and other variable objects. 
Here, we present the uninterrupted light curves for a sample of 64 cataclysmic variables (CVs) observed/discovered using the TMTS during its first three-year observations,
and we introduce new CVs and new light-variation periods (from known CVs) revealed through the TMTS observations.
Thanks to the high-cadence observations of TMTS, diverse light variations, including superhumps, quasi-periodic oscillations, large-amplitude orbital modulations, and rotational modulations, are able to be detected in our CV samples, providing key observational clues for understanding the fast-developing physical processes in various CVs. 
All of these short-timescale light-curve features help further classify the subtypes of CV systems.
We highlight the light-curve features observed in our CV sample and discuss further implications of minute-cadence light curves for CV identifications and classifications. 
Moreover, we examine the H$\alpha$ emission lines in the spectra from our nonmagnetic CV samples (i.e., dwarf novae and nova-like subclasses) and find that the distribution of H$\alpha$ emission strength shows significant differences between the sources with orbital periods above and below the period gap, which agrees with the trend seen from the SDSS nonmagnetic CV sample. }


\begin{document}

%%%%%%%%%%%%%%%%%%%%%%%%%%%%%%%%%%%%%%%%%%%%%%%%%%

%%%%%%%%%%%%%%%%% BODY OF PAPER %%%%%%%%%%%%%%%%%%

\section{Introduction}
%%Qichun, adding a review reference by Dan Maoz 2014 ARA&A for SN Ia progenitor
Cataclysmic variables (CVs) are semidetached binaries consisting of a white dwarf (WD) and a Roche-lobe-overflowing low-mass companion that is usually on or near the late-type main sequence. Observations of CVs provide opportunities to study accretion theories, the physics of compact objects, and~the evolution of WD binaries, as~some of them may ultimately end as Type Ia supernova explosions \citep{2014ARA&A..52..107M,2023RAA....23h2001L} or form an AM~CVn system with mHz gravitational waves (GWs) detected via space-born GW observatories \citep{Nelemans+etal+2001+AMCVn,2003MNRAS.340.1214P,Luo+etal+2016+TianQin,LISA+2017}. 
%are natural laboratories for accretion physics, so they attract extensive studies like X-ray binaries. Semi-detached binaries, consisting with a white dwarf (WD) and a low-mass companion, are called cataclysmic variables (CVs). 
%The donor in most cases is on or near the late-type main sequence, filling its Roche lobe. 

In CV systems, material from the donor star will be accreted onto the WD companion through the inner Lagrangian point $L_1$ \citep{1995cvs..book.....W} and form a surrounding accretion disc or fall into the magnetic poles of the WD, depending on the strength of the WD's magnetic field. 
Nonmagnetic CVs have two main subclasses, namely dwarf novae (DNe) and nova-like variables (NLs), which are both disk-dominant systems. The~dominant difference between them is that the DN subclass undergoes recurrent outbursts, while the NL subclass does \mbox{not \citep{2020MNRAS.495..637K}.}
In comparison, magnetic CVs can also be divided into two subclasses, intermediate polars (IPs, or~DQ~Her) with weak magnetic fields and polars (or AM~Her/AM) with stronger magnetic fields (e.g., 10--200~MG) \citep{2004MNRAS.351.1423B}. The~magnetic fields of IPs are not strong enough to prevent the formation of a disk, while their accretion mode may switch among different states (i.e., high/low state) \citep{2020MNRAS.495.4445K}.
Owing to the strong magnetic fields of polars, their accreted material follows the magnetic field lines to reach directly near the WD's magnetic poles and form accretion columns. 
The collision of subsonic falling flows against the WD photosphere leads to the formation of a shock, and~the shock-heated emission contributes significantly to the radiation of polars \citep{2000SSRv...93..611W,2015A&A...579A..25B}.

Superhumps are the periodic light-curve modulations of CVs with a photometric period comparable to their orbital period and an amplitude of about 0.3--0.4~mag \citep{1995cvs..book.....W}. 
The modulation period is slightly longer than the orbital period in positive superhumps (pSHs), while the modulation period is shorter than the orbital period in negative superhumps (nSHs) \citep{2023MNRAS.519..352B}. The~pSHs were proposed to be induced via tidal instability in the disk \citep{1996PASP..108...39O}. Tidal interaction with the secondary will impose an elliptical deformation to the \mbox{disc \citep{1991MNRAS.249...25W}} when it expands to the 3:1 resonance region during the outburst. The~precession of the eccentric accretion disk results in the light-curve modulations seen in those pSHs. In~contrast, the~nSHs are believed to be the retrograde precession of a tilted disk \citep{2009MNRAS.398.2110W, 2015ApJ...803...55T}.

%The CVs were divided into two groups by a period gap of 2--3~hr 
A prominent property of the CV population is the 2--3~H orbital period gap \citep{2006MNRAS.373..484K}.
%A well-known characteristic of CVs is the 2--3~hr gap . 
In binary evolution theories, angular-momentum loss (AML) can drive the orbital contraction of CVs. 
%, the orbits of binary systems generally contract with the evolution.
%binaries generally evolving from long orbital period to short orbital period. 
Magnetic braking is the dominant driving mechanism of AML for long-period CVs, leading to a typical mass-transfer rate of $\dot{M}\approx 10^{-9}$--$10^{-8}~{\rm M}_{\odot}~{\rm yr}^{-1}$ \citep{1983A&A...124..267S}, while GW radiation is the dominant AML mechanism for those short-period CVs, and it induces a mass-transfer rate of $\dot{M}\approx 5\times 10^{-11}~{\rm M}_{\odot}~{\rm yr}^{-1}$ \citep{1984ApJS...54..443P}.
%The main mechanism of angular momentum loss (AML) for long-period CVs is magnetic braking, with typical mass-transfer rates of $\dot{M}\approx 10^{-9}$--$10^{-8}~M_{\sun}~{\rm yr}^{-1}$ \citep{1983A&A...124..267S}. 
%For short-period CVs, gravitational radiation is a dominant mechanism of AML with $\dot{M}\approx 5\times 10^{-11}~M_{\sun}~{\rm yr}^{-1}$ \citep{1984ApJS...54..443P}. 

The spectra of CV systems usually show H$\alpha$ emission, which is thought to be related to the optically thin outer regions of the disks \citep{1980ApJ...235..939W}. High-inclination CV systems even exhibit obvious double-peaked profiles due to Doppler broadening. 
However, during outburst, DNe tend to show narrow Balmer emission lines with broad absorption wings, which can be explained by a hot, optically thick disk with a relatively cool and optically thin outer region \citep{1989ApJ...337..432C}. 
\citet{2006PASA...23..106S} adopted a discriminant function and principal-component analysis to study the ratios of Balmer emission lines, and~they found that the discriminant function can separate DNe from other subclasses. 
%Inspired by that work, we make an attempt to investigate the line-formation process of H$\alpha$ on the two sides of period gap.

%CVs are natural laboratories for accretion physics, so they attract extensive studies like X-ray binaries. 
% 
Based on the well-sampled light curves from several telescopes like \textit{Kepler} \citep{2010ApJ...713L..79K} and the {\it Transiting %MDPI: Please confirm if the italics is unnecessary and can be removed. Please check the whole paper.
 Exoplanet Survey Satellite} (\textit{TESS}; \citet{2015JATIS...1a4003R}), light variations intrinsic to CVs have been examined \citep{2016MNRAS.460.2526O, 2022MNRAS.509.4669B, 2023MNRAS.519..352B}, which helps improve {our understanding} of accretion instability. TMTS can also provide well-sampled light curves for CVs; it is a multitube telescope system consisting of four 40~cm optical telescopes, and it has a field of view (FoV) of up to about $18~{\rm deg}^2$ \citep{2020PASP..132l5001Z}. This facility has discovered/monitored over 1100 short-period variable sources during the first two-year monitoring of the Large Sky Area Multi-Object Fiber Spectroscopic Telescope (LAMOST; \citet{2012RAA....12.1197C,2012RAA....12..723Z}) skyfields, including eclipsing binaries, pulsating stars, cataclysmic variables, and~so on \citep{2022MNRAS.509.2362L, 2023MNRAS.523.2193L, lin2023sevenearthradius, 2023NatAs...7..223L, 2023MNRAS.523.2172L, 2024MNRAS.528.6997G}. This paper aims to release the light curves of CVs and the candidates identified in the first 3~yr, and~it presents the analysis of their periodicities, light-curve features, and~spectroscopic~properties.

\section{Observations and Data~Analysis}

%%Qichun, have you included the 3rd-year data? if so, the total plates of LAMOST and number of uninterrupted light curves should be revised
\subsection{TMTS CV~Sample}
\label{selection}
In the first 3~yr survey, TMTS covered a total of $6977~{\rm deg^2}$ (449 LAMOST/TMTS plates with more than
100 visits), and~it produced 19,099,266 uninterrupted light curves with at least 100 valid measurements for about 20 million sources. After~cross-matching the CV catalog that contains 5478 CVs and CV candidates from the {SIMBAD \mbox{database \citep{2000A&AS..143....9W}}} and other works in the literature \citep{2001PASP..113..764D, 2003A&A...404..301R, 2011AJ....142..181S, 2021ApJS..257...65S}, we obtained the light curves for a total of 57 CVs and 5 CV candidates. Additionally, 
this work includes two new CV candidates that were first discovered through \mbox{TMTS---TMTS} J04405040+6820355 and TMTS J06183036+5105550.
%following the process described in \citep{2023MNRAS.523.2172L}.
Among the 64~CV samples, 57 have been included in the International Variable Star Index (VSX) \citep{2006SASS...25...47W}, with~classifications recorded as 29 DNe, 12 NLs, 11 IPs, and~5~AMs.

Note that most of the samples were uninterruptedly observed within single nights except for five sources that were monitored on two separate nights.  
The procedures of TMTS photometry and calibration are described by \citet{2022MNRAS.509.2362L}. 
Uninterrupted light curves of the 64 CVs are shown in \cref{fig:lc,fig:lc3,fig:lc4,fig:lc5,fig:lc6}. These light curves typically span 4--12~h, with~a cadence of about 1~min. 
For those known CVs, we also labeled the \mbox{IAU-sanctioned names.}
%Some of our identifications are well-studied objects, like TMTS J075653+0858318 (SDSS 0756+0858; \citealt{2014AJ....147...68T, 2015A&A...573A..61S, 2015AJ....149..128T}).

%%Qichun, the total CV sample is 63, what are about the spectra for the other 14 CVs? need to be addressed in the following paragraph.
%The other CVs have no spectra, some of them (maybe 3 sources, I will check it later) have lick spectra that have not added yet. I will add them tomorrow. 

%%Qichun, in Figure~2, you mentioned that spectra are available for 59 sources, which is inconsistent with the number indicated below, please clarify...
%Dear Prof., that is a mistake, I correct it.
Benefiting from our observation strategy, among~the 64 samples, spectra of 27 are available from the LAMOST DR7. 
We obtained spectra of an additional 29 objects with the Xinglong 2.16~m telescope (XLT) and the Lick 3~m Shane telescope (LST). 
Nevertheless, spectroscopic data are still absent for seven objects of our collected samples. Figure~\ref{fig:spec} shows the spectra of our 57 CV samples.
%These CV spectra typically exhibit broad Balmer emission lines, while the He emission lines are landmarks of NLs and magnetic CVs \citep{2020AJ....159...43H,2021ApJS..257...65S}. 
%, such as TMTS J21332987+5127219, TMTS J22180793+5613115, and TMTS J22362065+6622063. 
%For the 63 sources, 22 have spectra from Xinglong 2.16 m telescope, and 24 have LAMOST spectra.  
%The spectra of 4 candidates are also shown. 
Here, we give brief introductions to some candidates listed in Table~\ref{tab1}. TMTS J03394099+4148057 was identified as an NL candidate \citep{2020AJ....159...43H} with a radial-velocity period of 3.54~h \citep{2020AJ....160..151T}.
Our spectrum of this object revealed the presence of prominent H$\alpha$ $\lambda6563$ and He~I $\lambda6678$ and %Please check that the intended meaning has been retained.
$\lambda7065$ emission lines.
 %We observed this source with LST, {further confirming} the emission lines of H$\alpha$  and He I .
TMTS J09011350+1447046 is a CV candidate that was first discovered by \citet{2009AJ....137.4011S}, and~its LAMOST spectrum exhibits a broad, moderately strong H$\alpha$ emission \citep{2021ApJS..257...65S}. TMTS J07200739+4516113 was initially discovered by \citet{2018ATel11626....1D}, and it has been suggested to be polar, according to the helium emission features seen in the LAMOST spectra \citep{2021ApJS..257...65S} or its large amplitude of light variation \citep{2023MNRAS.523.2172L}.  Owing to the lack of detailed analysis and constraints on the accretion structure, we tentatively treat it as a CV candidate in this work. 

\begin{table}[H]
\caption{Catalog %MDPI: we moved all tables and figures after the first citation, please confirm.
 of cataclysmic variables from the TMTS~observations. \label{tab1}}
	\begin{adjustwidth}{-\extralength}{0cm}
%	\begin{tabular}{@{}lllllll}
	\begin{tabularx}{\fulllength}{lllllll}
			\toprule
			\textbf{Name}  & \textbf{Start Time}  & \boldmath{$P_{\rm pho}$} & \textbf{Amplitude} & \textbf{Feature} &\textbf{VSX Name} &  \textbf{Reference} \\
		 & \textbf{(MJD)}  & \textbf{(min)} & \textbf{(mag)} & && \\  
			\midrule
    & &  Dwarf novae&  &  &   \\
    \midrule
TMTS J00060995+5558501& 59886.42805 %MDPI: please confirm if the format of time is correct.
&&&&FI Cas & 1 \\
TMTS J01010887+4323259& $59877.42796$&&&&IW And & 1\\
TMTS J01043552+4117576&  $59877.44097$&$145.0 \pm 0.3$~~~~~~~~~~~~~~~~~~~~~&$0.165$&L,R&RX And& 1\\
TMTS J01101317+6004349&  $59509.46110$&&&&HT Cas & 1 \\
TMTS J01153217+3737354&  $59876.43408$&&&&FO And & 1\\
TMTS J01275052+3808122&  $59876.47894$&&&&1RXS J0127+38~~~~~~~~~~~~~ & 1\\
TMTS J02135093+5822527&  $59596.43877$&&&&TZ Per& 1\\
TMTS J02262311+7118314& $59210.41097$&&&&AM Cas & 1 \\
\bottomrule
\end{tabularx}
 \end{adjustwidth}
 \end{table}

\begin{table}[H]\ContinuedFloat
\small
\caption{{\em Cont.}}
 \begin{adjustwidth}{-\extralength}{0cm}
%	\begin{tabular}{@{}lllllll}
		\begin{tabularx}{\fulllength}{lllllll}
			\toprule
			\textbf{Name}  & \textbf{Start Time}  & \boldmath{$P_{\rm pho}$} & \textbf{Amplitude} & \textbf{Feature} &\textbf{VSX Name} &  \textbf{Reference} \\
		 & \textbf{(MJD)}  & \textbf{(min)} & \textbf{(mag)} & && \\  
			\midrule
    & &  Dwarf novae&  &  &   \\
    \midrule
TMTS J02500008+3739219& $59194.43101$&&&&PY Per& 1\\
TMTS J03124571+3042477&  $59199.42001$&&&&CRTS J0312+30 & 2\\
TMTS J03321548+5847219 {*}&  $59899.67009$&&&&AF Cam& 1\\
-&  $59207.41269$&&&&- & -\\

TMTS J04023898+4250447 {*}&  $59548.42925$&&&&V1024 Per & 3\\
-& $59548.42836$&&&&-&-\\
TMTS J04083502+5114484&  $59943.41288$&$208.00 \pm 1.45$&$1.344$&H&FO Per & 1\\
TMTS J04184443+5107313 {*}&  $58877.44968$&&&&NS Per & 1\\
-&  $59943.41273$&$110.98 \pm 1.33$&$0.177$&L&- & -\\
TMTS J04260927+3541442&  $59554.42448$&&&&MASTER OT J0426+35 & 4\\
TMTS J04463363+4857559&  $58852.42751$&&&&ASASSN-15rs & 5\\
TMTS J05235177+0100303&  $59249.43854$&$135.83 \pm 3.39$&$0.213$&L&BI Ori & 1\\
TMTS J05581781+6753456&  $59598.45463$&$46.4 \pm 0.4$&$0.300$&L&LU Cam & 1\\
TMTS J06132238+4744248&  $59202.41699$&&&&SS Aur & 1\\
TMTS J07485955+3125121&  $59281.46741$&$82.1 \pm 0.8$&$0.429$&E&SDSS J0748+31 & 5\\
TMTS J08442711+1252322 {*}&  $59204.65922$&&&E&AC Cnc & 1\\
-&  $58898.47159$&&&E&- &1\\
TMTS J08534425+5748402&  $59216.69161$&&&&BZ UMa & 1\\
TMTS J09121621+5053531&  $59617.68575$&$77.9 \pm 0.9$&$0.339$&L & DI UMa & 1\\
TMTS J10020745+3351005& $59941.67607$&&&R&RU LMi & 1\\
TMTS J10043481+6629148&  $59685.52063$&$133.76 \pm 2.06$&$0.167$&L&LN UMa & 1\\
TMTS J10202651+5304330&  $59631.73682$&&&&KS UMa& 1\\
TMTS J10543054+3006090&  $59248.54145$&&&&SX LMi & 1\\
TMTS J10565691+4941183&  $59672.48447$&$193.12 \pm 1.55$&$0.295$&L&CY UMa &1\\
TMTS J12393204+2108063&  $59676.48460$&&&E&IR Com &1\\
\midrule
    &  &   Intermediate polars&  &  & \\
  \midrule
 TMTS J00225764+6141076&  $59507.45075$&$9.372 \pm 0.005$&$0.239$&R,L&V1033 Cas &1\\
TMTS J00284893+5917207&  $59181.43487$&&&&{V709 Cas}&1\\
TMTS J00551974+4612566 {*}&  $59873.51739$&&&&{V515 And} & 6\\
-&  $59869.43510$&$203.1 \pm 0.9$&$0.112$&L&- & -\\
TMTS J03311195+4354154& $59530.42598$&&&&GK Per &1\\
TMTS J05474838+2835104&  $59235.42950$&&&&FS Aur& 1\\
TMTS J06251631+7334386&  $59597.42651$&$19.785 \pm 0.009$&$0.209$&R,L&MU Cam & 1\\
TMTS J06274641+0148100&  $59250.44675$&&&E&{V902 Mon} &7\\
TMTS J07511729+1444239&  $59192.66469$&&&&PQ Gem& 1\\
TMTS J08382201+4838023&  $58907.51608$&&&&EI UMa & 1\\
TMTS J21334362+5107248&  $59835.54625$&&&&1RXS J2133+51 & 1\\
TMTS J22165027+4646412&  $59822.49572$&&&E&HBHA 4705-03 & 8\\
\midrule
    &  &  Nova-like variables&  &  &  \\
 \midrule
TMTS J01385585+2429393&  $59931.42144$&&&&SDSS J0138+24 &9\\
TMTS J05064797+8319233&  $59155.69393$&$85.8 \pm 0.3$&$0.327$&E&V1024 Cep & 1\\
TMTS J05572400+7241528&  $59597.42662$&$188.40 \pm 2.08$&$0.265$&L&LS Cam& 1\\
TMTS J06293373+7104361&  $59597.42655$&&&&BZ Cam & 1\\
TMTS J07565314+0858318&  $59636.46295$&$95.0 \pm 0.4$&$0.300$&E&SDSS J0756+08 & 10 \\
TMTS J08021533+4010463&  $58863.60606$&$55.8 \pm 0.5$&$0.090$&L&SDSS J0802+40 & 1\\
TMTS J08125687+1911572&  $59640.46993$&&&E&NS Cnc & 1\\
TMTS J08223605+5105242&  $59182.64784$&&&E&BH Lyn & 1\\
TMTS J09030895+4117467&  $58941.48244$&&&&BP Lyn &1\\
TMTS J09201115+3356421& $59207.67148$&$110.3 \pm 0.4$&$0.125$&L&BK Lyn & 1\\
TMTS J10481806+5218295&  $59665.50191$&$191.52 \pm 1.18$&$0.146$&L&LY UMa & 1\\
TMTS J23400423+3017476&  $59846.49587$&$122.4 \pm 0.2$&$0.151$&L&{V378 Peg} &1\\
			\bottomrule
		\end{tabularx}
	\end{adjustwidth}
	%\noindent{\footnotesize{* Tables may have a footer.}}
\end{table}
\unskip

\begin{table}[H]\ContinuedFloat
\small
\caption{{\em Cont.}}
	\begin{adjustwidth}{-\extralength}{0cm}
%	\begin{tabular}[l]{@{}lllllll}
		\begin{tabularx}{\fulllength}{lllllll}
		\toprule
			\textbf{Name}  & \textbf{Start Time}  & \boldmath{$P_{\rm pho}$} & \textbf{Amplitude} & \textbf{Feature} &\textbf{VSX Name} &  \textbf{Reference} \\
		 & \textbf{(MJD)}  & \textbf{(min)} & \textbf{(mag)} & && \\  
			\midrule
   &  &   Polars&  & &  \\
  \midrule
TMTS J00185684+3454451&  $59148.44826$&&&&{V479 And} & 1\\
TMTS J05154141+0104402& $59249.44113$&&&E&V1309 Ori & 1\\
TMTS J07112595+4404048& $59190.65350$&$118.56 \pm 0.13$~~~~~~~~~~~~~~~~~&$1.743$&E&{V808 Aur}& 11\\
TMTS J11042565+4503131& $59673.50834$&&&E&AN UMa & 1\\
TMTS J13075377+5351303&  $58923.56208$&$79.6 \pm 0.2$&$0.681$&H&EV UMa & 1\\
\midrule
   &  &   Candidates&  &   \\
  \midrule
TMTS J02461608+6217029& $59529.43927$&$109.7 \pm 0.4$&$0.156$&L&{V495 Cas}& new\\
TMTS J03394099+4148057&  $59544.42431$&$189.5 \pm 0.7$&$0.212$&L& None & 9\\
TMTS J03471387+1611083&  $59883.58651$&$97.7 \pm 0.8$&$0.262$&L&MLS\_J0347+16  & new\\
TMTS J04405040+6820355&  $59927.66419$&$110.0 \pm 1.1$&$0.262$&L&None  & new\\
TMTS J06183036+5105550& $59202.41861$&$107.4 \pm 0.4$&$0.139$& L&None  & 12\\
TMTS J07200739+4516113&  $59190.67704$&$90.9 \pm 1.0$&$>1$&H&MASTER OT J0720+45 & 13\\
TMTS J09011350+1447046&  $59209.66682$&$89.57 \pm 1.31$&$0.076$&L&SDSS J0901+14 & 14\\
			\bottomrule
		\end{tabularx}
	\end{adjustwidth}
\noindent{\footnotesize{{Note%MDPI: We removed the italics. Please confirm this revision.
}: Column (1), TMTS designation; column (2), the~time when the TMTS observation started; column (3), photometric period corresponding to the maximum power in the LSP of the TMTS observation; column (4), peak-to-peak amplitude obtained from the best-fitting model of fourth-order Fourier series; column (5), light-curve feature defined in Section~\ref{lc_feature} (E, Eclipse; R, Rapid periodic variation; L, Low-amplitude periodic variation; H, High-amplitude periodic variation); column (6), name in VSX; column (7), reference of the corresponding source. The~* symbol indicates that the source was observed on two separate~nights.
          References: (1) \citet{2001PASP..113..764D}; (2) \citet{2014ApJS..213....9D}; (3) \citet{refId0}; (4) \citet{2012ATel.4441....1D}; (5) \citet{2016PASJ...68...65K}; (6) \citet{2012MNRAS.422.1518K};
          (7) \citet{2007MNRAS.382.1158W};        (8) \citet{2013AstL...39...38Y};
           (9) \citet{2020AJ....159...43H};
          (10) \citet{2011AJ....142..181S};
          (11) \citet{2010IBVS.5923....1T};
          (12)~\citet{2022MNRAS.509.2362L};
          (13) \citet{2018ATel11626....1D};
          (14) \citet{2009AJ....137.4011S}.}}
%     \end{tablenotes}
	%\noindent{\footnotesize{* Tables may have a footer.}}
\end{table}
\unskip

\subsection{Light-Curve~Analysis}
\label{lightcurve}

\subsubsection{Periodic~Variations}

We searched for periodic variations in the
TMTS light curves using the Lomb--Scargle periodogram (LSP; \citet{1976Ap&SS..39..447L, 1982ApJ...263..835S}). The~frequency range was set to $2/T \leq f \leq f_{\rm nyq}$, where $T$ is the time span of each light curve, and $f_{\rm nyp}$ is the Nyquist frequency, estimated as half of the average sampling rate. 
%The frequency step is $\Delta f=1/n_0T$, where $n_0$ represents the oversampling factor to account for irregular sampling \citep{2021MNRAS.505.2954C} and is adopted as 20 in our analysis. 
The method adopted to compute the LSP for the TMTS light curves was already described in detail by \citet{2022MNRAS.509.2362L}. The~false-alarm probability (FAP) was estimated as
\begin{equation}
    {\rm FAP}=1-[1-{\rm exp}(-z)]^{N_{\rm eff}}\, ,
	\label{eq:FAP}
\end{equation}
where $z$ is the LSP power, and $N_{\rm eff}$ is the effective number of independent frequencies. $N_{\rm eff}$ is approximated as $f_{\rm nyp}T$ \citep{2018ApJS..236...16V}. 
We adopted a significance threshold of FAP = 0.001 throughout this~work.
 
%The frequency corresponding to the maximum LSP power, $f_{\rm pho}$, are reported in Table~\ref{tab1}. 
From the TMTS LSPs of CV samples, we determined the photometric period corresponding to the maximum LSP power, $P_{\rm pho}$, if~the maximum LSP power was higher than the threshold (see Table~\ref{tab1}).
Then, we used a compound model of fourth-order Fourier series with a period equal to $P_{\rm pho}$ plus a second-order polynomial to fit all TMTS light curves (see Equation~(7) of \citet{2022MNRAS.509.2362L}). 
The peak-to-peak amplitudes obtained from the fourth-order Fourier series are listed in Table~\ref{tab1}.
%The peak-to-peak amplitude  of fourth-order Fourier series are taken as the amplitude of the light curve.
%We selected only the $P_{\rm pho}$ smaller than $T/2$ with power exceeding the threshold to avoid possible strong power caused by red-noise pollution, owing to the relatively short temporal coverage of the TMTS observations. As a result, some sources in Table~\ref{tab1} do not have $P_{\rm pho}$ and relevant amplitudes listed.

\subsubsection{Light-Curve~Features}
\label{lc_feature}
\begin{figure}[H]
	\includegraphics[width=10.5 cm]{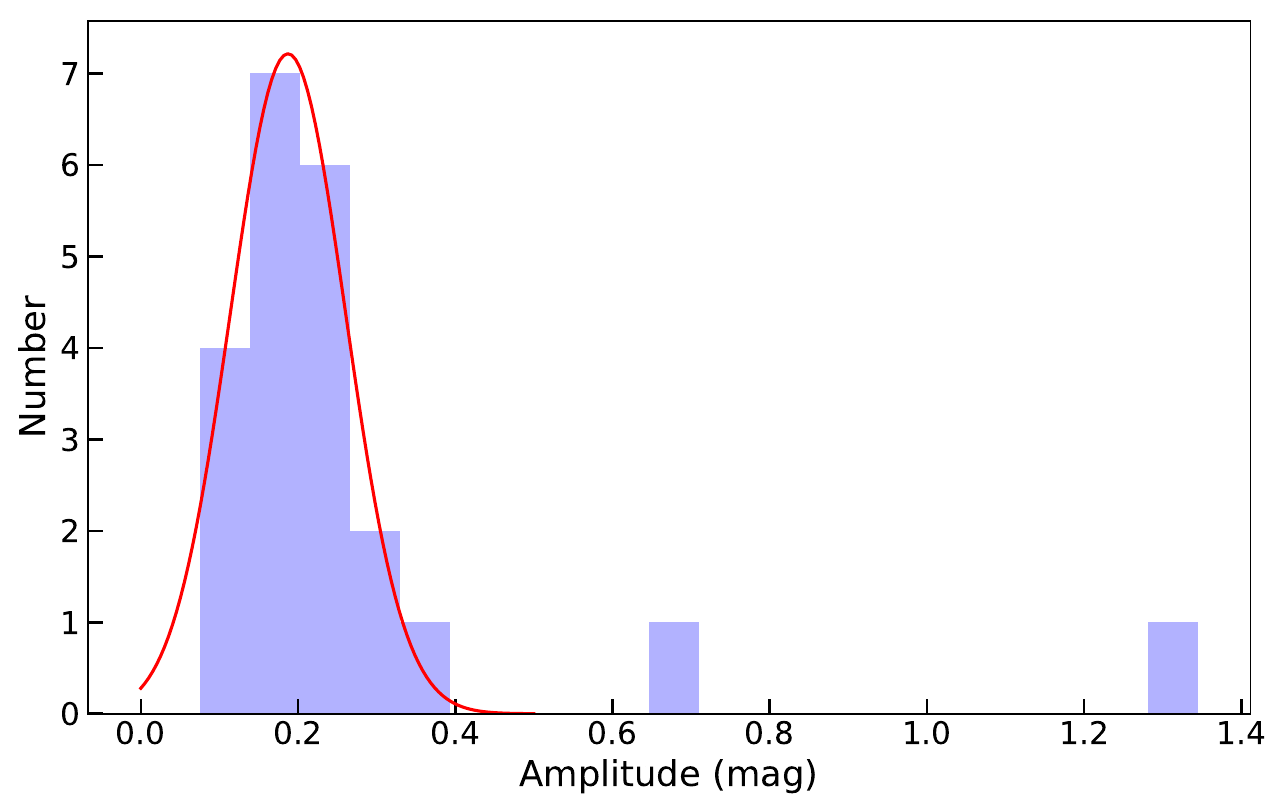}
    \caption{Distribution of the amplitude of periodic variations from noneclipse CVs. The~red line represents a Gaussian~fit.}
    \label{fig:amplitude}
\end{figure}

With the minute-cadence photometry from TMTS, we can characterize the uninterrupted light curves of these CVs. 
Here, all short-timescale light-curve features were classified into four distinguishable types, as~follows.
\begin{itemize} 
    \item \textbf{Eclipse.%MDPI: please confirm if the bold in list is necessary. the same below.
} Thanks to densely sampled photometry, the~TMTS light curves allow us to characterize the detailed profiles of eclipses for the eclipsing CVs. In~our CV samples, the~eclipse depth ranges from a few tenths of magnitude to more than 2.0~mag.
    The emergence of an eclipse provides direct evidence  supporting the notion that the CV system has an orbital inclination approaching $90^\circ$.
    \item \textbf{Low-amplitude periodic variation}.
    With the amplitude distribution of noneclipse CV systems (see Figure~\ref{fig:amplitude}), 20 CV samples presented  periodic modulation amplitude of about 0.1--0.4 mag, while 2 CV samples exhibited a significantly higher modulation amplitude. Low-amplitude periodic variations here are defined as periodic variations with an amplitude lower than 0.4 mag, which are typically caused by the hump/superhump of CVs and are thus tightly related to their orbital periods.
    \item \textbf{High-amplitude periodic variation}.
    In contrast, high-amplitude periodic variations represent modulations with an abnormally high amplitude. Here, we roughly define them as periodic variations with an amplitude larger than 0.4 mag.
    \item \textbf{Rapid periodic variation}. Since the quasi-periodic oscillations of DNe and rotation modulations of IPs are difficult to  distinguish with the single-night light curves, here, the rapid periodic variations represent all periodic or quasi-periodic variations below 20 min, significantly shorter than the periodic variations induced via \mbox{orbital~modulations.}

\end{itemize}

\vspace{-6pt}

\subsection{Spectroscopic~Analysis}
\label{spec}

With the spectra collected from LAMOST, LST, and~XLT, we further performed a spectroscopic study of our CV samples. ({Note %MDPI: Footnotes are not allowed in this journal. We have moved the content of the footnote to the corresponding position of the mark, please confirm
 that the CV candidates are not included in this study.}) We selected those CVs with better determinations of subtypes and orbital periods from our 64 CV samples, excluding 7 candidates and 3 sources without an orbital period.
%Then we downloaded the spectra of LAMOST and SDSS CVs to enrich our sample. 
Only spectra with a signal-to-noise ratio (SNR) %Please check that the intended meaning has been retained.
 $>10$ and variation coefficient (VC) %Please check that the intended meaning has been retained.
  $<1$ are included in our analysis, where VC is defined as the ratio of the standard deviation to the mean value of the SNR. VC is used to evaluate the quality of the spectrum, which is acceptable when VC $< 1$ \citep{2021ApJS..253...51L}. In~addition, only those spectra showing emission features were included in the analysis since we are interested in the emitting process in the CVs. Finally, 44 CVs were included in the spectroscopic analysis.
%{to focus on emitting process}, while those with only H$\alpha$ absorption lines are discarded from the analysis. 

%%Qichun, it is not clear what you meant for "a correction" below, please specify it; why you state that "As most values of ${\chi_{\rm red}^2}$ are less than 5, the best-fit values of ${\rm EW}_{\rm H\alpha}$ should be reliable."?  As a reasonable reduced chi-square should be about 1.0.

%Dear Prof., yes, for physical models the best solution is assumed to be found when reduced chi square is close to 1.0, but when the reduced chi square is small like 5, the solution is already fairely well.
The equivalent width (EW) of H$\alpha$ emission, ${\rm EW}_{{\rm H}\alpha}$, was calculated by integrating the flux excesses above the continuum:
\begin{equation}
    {\rm EW}=\int{\frac{F_{\lambda}-F_c}{F_c} d\lambda}\, ,
	\label{eq:ew}
\end{equation}
where $F_\lambda$ represents the emission flux at wavelength $\lambda$, and $F_c$ denotes the continuum flux. 
$F_\lambda$ is the flux in the wavelength range of 6480--6640~\AA. Such a wide range can include the entire broadened profile of H$\alpha$ emission. 
The continuum flux was obtained by linearly fitting the spectral ranges 6480--6510~\AA\ and 6620--6640~\AA, and~no emission component emerged over these ranges in our spectra.
%The EW of H${\alpha}$ is obtained by integrating the normalized excesses above the continuum ($F_{\lambda}-F_c/F_c$), dubbed as $F_{\rm sub}(\lambda)$), while 
Additionally, the~full width at half-maximum intensity (FWHM) is the wavelength interval between halves of the maximum flux in the emission~lines. 

We used Monte Carlo simulations to estimate the uncertainties of ${\rm EW}_{{\rm H}\alpha}$ and ${\rm FWHM}_{{\rm H}\alpha}$. The~flux of the spectra was randomly sampled according to the uncertainty at each wavelength. After~repeating the simulation 100 times, the~standard deviations of the measured ${\rm EW}_{{\rm H}\alpha}$ and ${\rm FWHM}_{{\rm H}\alpha}$ were taken as the corresponding uncertainties. The~typical uncertainties for the two quantities are both $\sim$1~\AA.

\section{Individual~Systems}
\label{individual}

The minute-cadence observations  from TMTS enabled us to study the short-timescale light variations of all these CV samples, including the newly discovered ones (see \mbox{Table~\ref{tab1}).} 
%Benefiting from densely sampling rates, 
In particular, the~TMTS data reveal new photometric periods for some known CVs, together with the observations from  {\textit{TESS}} and the Zwicky Transient Facility (ZTF; {\citet{2019PASP..131a8002B, 2019PASP..131a8003M})}.
%, we analyzed these new CVs and photometric periods in details.

%For the CVs/CV candidates in our sample, we inspected their light curves carefully. 
%Since TMTS is a minute-cadence survey, it is expected that it can observe short-time scale variables detailedly. We report these new findings in this section, and for some cases in order to investigate the origin of periods in TMTS light curves, we seek a hand to the Transiting Exoplanet Survey Satellite (\textit{TESS}; \citet{2015JATIS...1a4003R}) observations from the Mikulski Archive for Space Telescopes at the Space Telescope Science Institute. }

\subsection{Newly Discovered CVs and CV~Candidates}

\subsubsection{TMTS J04405040+6820355}
\label{new cv}

\begin{figure}[H]
% \begin{adjustwidth}{-\extralength}{0cm}
% \centering
	\includegraphics[width=13.5 cm]{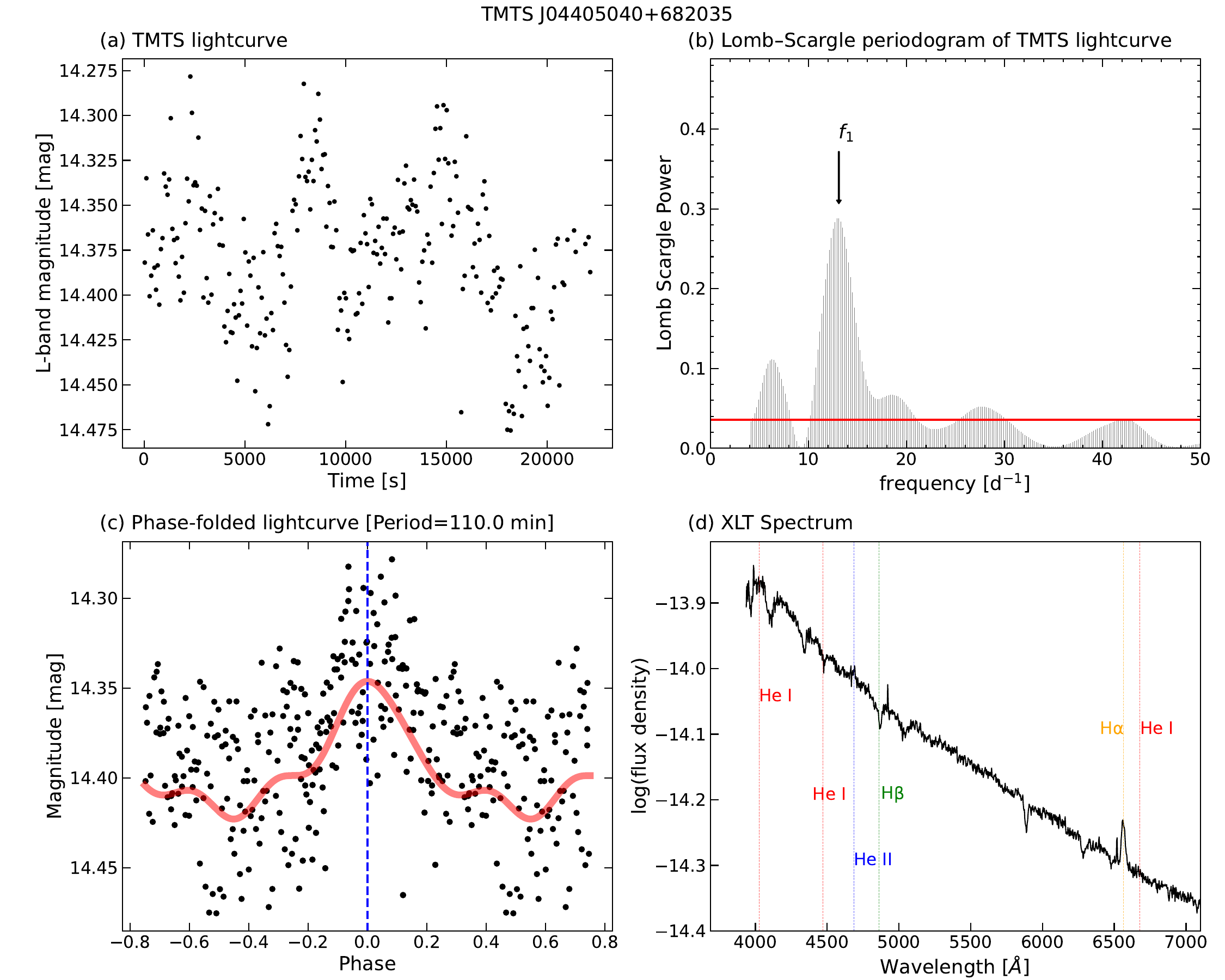}
% \end{adjustwidth}
    \caption{(\textbf{a}) TMTS %MDPI: 1 Please use commas to separate thousands for numbers with five or more digits (not four digits) in the picture, e.g., "10000" should be "10,000". 2 please confirm if red line in figure need further explanation.
 light curve of TMTS J04405040+6820355; (\textbf{b}) Lomb--Scargle periodogram of TMTS J04405040+6820355; (\textbf{c}) phase-folded TMTS light curve with $P_1=110.0$ min, with~the red line representing the best-fit model of fourth-order Fourier series; (\textbf{d}) the XLT spectrum of TMTS J04405040+6820355.}
    \label{fig:04405040+6820355}
\end{figure}

%TMTS J04405040+6820355 (hereafter J0440) is a new CV candidate discovered by TMTS.
TMTS J04405040+6820355 (hereafter J0440) shows a significant light-variation period of $P_1=110.0\pm 1.1$~min in the TMTS light curve obtained on  14 December 2022 (UTC dates are used throughout this paper); see Figure~\ref{fig:04405040+6820355}. Broad H$\alpha$ emission is  superimposed on a blue continuum in the spectrum obtained with the Xinglong 2.16~m telescope on 2 October 2023 (Figure~\ref{fig:04405040+6820355}). Given the location of J0440 in the Gaia color-magnitude diagram ($M_{\rm abs, G}=5.4$~mag and $G_{\rm Bp-Rp}=0.13$~mag), we infer that J0440 is a new CV candidate (see also \citet{2023MNRAS.523.2172L}).

%The absence of He \Rmnum{2} $\lambda$4686 line in the spectrum indicates that {J0440 is a DN rather a NL \citep{2020AJ....159...43H}}. %注意一定要加引文，并且确认 是不是缺少He线就说明这个源是DN。
The five-year light curves provided via ZTF failed to reveal any outburst for this source; {along with the presence of H$\beta$ absorption, this implies that it might be an NL with an optically thick disk.}

%implying that it could have a very long-duration quiescence like those WZ Sge variables.

%with the reddening-removed G band absolute magnitude $M_{\rm abs, G}=5.4$ mag and Gaia color $G_{\rm Bp-Rp}=0.13$ mag from Gaia DR2 database \citep{2018A&A...616A...1G}, located below the main-sequence. 
%The spectrum taken with the Xinglong 2.16~m telescope observed in JD 2460220.3396 shows wide H$\alpha$ emission superimposed on a blue continuum, typical of normal CVs (see Figure~\ref{fig:spec}). 
%We analysed the H$\alpha$ line with the process described in Section~\ref{spec}, and found that the FWHM is 28.16~\AA. 
%{The spectrum has no He lines, so we speculate it is a DN. From ZTF 5-year light curve we do not see any outburst, a long-term photometric monitor for this source is required. }
%From the TMTS light curve we found a period $P_1=110.0\pm 1.1$ min with strong power, which is likely a orbital period. The LSP periodogram and phase-folded light curve are shown in Figure~\ref{fig:04405040+6820355}.

\subsubsection{TMTS J06183036+5105550}
\label{J0618}

TMTS J06183036+5105550 (hereafter J0618) is a new CV discovered via TMTS and  reported in the catalog of TMTS short-period variable stars \citep{2023MNRAS.523.2172L}. Its spectrum shows a faint He~II emission line around 4686~\AA. According to the \textit{Swift}/XRT observation conducted on  29 October 2022, this source has an X-ray luminosity of $\sim$$10^{31}~ {\rm erg~s^{-1}}$. Two periodicities, $P_1=107.4\pm 0.4$ min and $P_2=11.165\pm 0.004$ min, are revealed via the TMTS light curve. These clues favor the classification of this object as an intermediate polar~candidate.  

\begin{figure}[H]
	\includegraphics[width=10.5 cm]{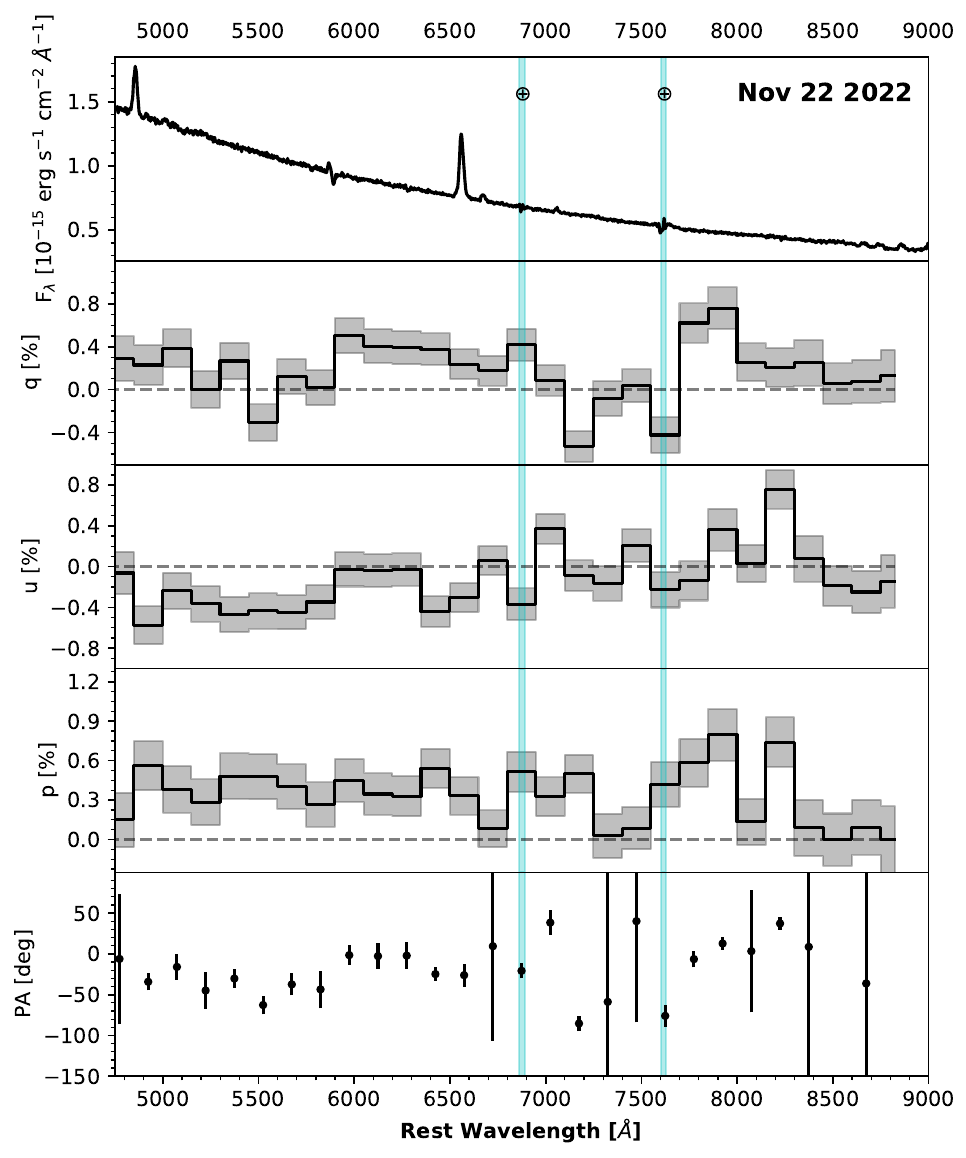}
    \caption{Spectropolarimetry %MDPI: In order to avoid a large amount of blank space on the page, we moved Figure 3 here. Please confirm.
 of TMTS J06183036+5105550 obtained with the Lick/Shane 3~m telescope on  22 November 2022. The~cyan vertical bands represent the regions of telluric correction. {The top panel presents the total-flux spectrum.} The panels below the total-flux spectrum represent the polarimetry after the ISP correction. The~gray-shaded area indicates the associated 1$\sigma$ uncertainty.}
    \label{fig:tmts_polarization}
\end{figure}

We further explored the properties of this source by conducting a polarization observation on  22 November 2022 with the Kast double spectrograph on the Shane 3~m telescope at Lick Observatory, USA. The~spectropolarimetric results are shown in Figure~\ref{fig:tmts_polarization}, where $q=Q/I$ and $u=U/I$ are the normalized Stokes parameters. $Q$ and $U$ describe the differences of fluxes when the electric vector oscillates in two perpendicular directions, and~$I$ is the total flux. The~observed polarization, $p$, is calculated as $p_{\rm obs}=\sqrt{q^2 + u^2}$, and~the polarization angle is calculated through ${\rm PA_{\rm obs}}=(1/2) \arctan (u/q)$.
%Note that the final reported values of $p$ and ${\rm PA}$ are derived through a debiasing procedure.  We corrected the polarization by polarimetric calibration to remove instrumental polarization. 
Since electromagnetic waves passing through interstellar dust will become polarized, the~galactic interstellar polarization (ISP) is also taken into account when computing the intrinsic polarization from the CV system.
%when deriving the intrinsic polarization because light passing through interstellar dust will become polarized. 
Following the procedure described by \citet{2022MNRAS.509.4058P}, the~ISP of J0618 was removed by subtracting the observed polarization from the intrinsically unpolarized star within the $1^{\circ}$ of this object.
%we measured the polarization of an intrinsically unpolarized star within $1^{\circ}$ of TMTS J06183036+5105550, which is then subtracted from the observed polarization of this CV system. 

As seen in Figure~\ref{fig:tmts_polarization}, a~linear polarization of $\sim$0.6\% was detected for J0618, comparable to the $0.8\%$ linear polarization reported for intermediate polar RE~0751+14 in the $R$ \mbox{band \citep{1993ApJ...410L.107P}.} 
The values of the polarization angle are noisy for some points, but~in general, they are not randomly scattered. 
In the case of magnetic CVs, material from the companion star is accreted onto the magnetic poles of the WD, leading to cyclotron radiation \citep{2009A&A...496..891B}. The~higher harmonics of the fundamental frequency will generate linear polarization \citep{1995cvs..book.....W}. Unlike polars, polarized light from IPs will be diluted due to the emission from the accretion disk, WD, and~so on. This leads to the fact that polarized emission is detected only in a few IP systems \citep{2012MNRAS.420.2596P}. 
With all the above observational evidence, we suggest that J0618 is an intermediate~polar.

\subsection{New Light-Variation Features from Known~CVs}
\unskip
\subsubsection{SDSS J013855.86+242939.2}
%%Qichun, no explanation for "some flicking seen in the light curve"...
SDSS J013855.86+242939.2 (TMTS J01385585+2429393, hereafter J0138) is a relatively poorly studied CV identified by cross-matching SDSS WD candidates with LAMOST spectra \citep{2015MNRAS.452..765G}. \citet{2020AJ....159...43H} classified it as a magnetic CV according to the comparable strength of H$\beta$ and He~II $\lambda4686$ lines. Nevertheless, they also emphasized that this classification was based only on the characteristic lines and may not be accurate. From~ZTF {\it r}-band observations, the~brightness of this CV can change from $\sim$15 mag to $\sim$18 mag. The~brightness variation of this CV resembles that of magnetic CVs \citep{2005AJ....130.2852K, 2022ApJ...928..164C}.
%the VY~Scl subclass of NLs \citep{2022MNRAS.516.2775S}. 

\begin{figure}[H]
  \begin{adjustwidth}{-\extralength}{0cm}
  \centering
	\includegraphics[width=15.5 cm]{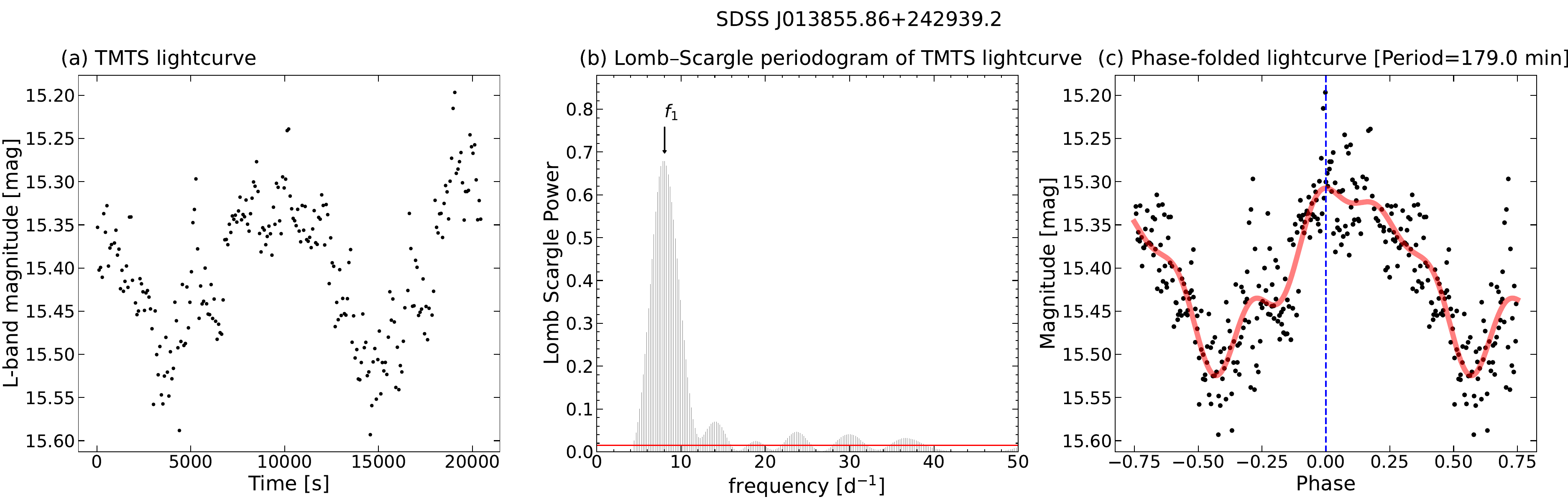}
 \end{adjustwidth}
    \caption{(\textbf{a}) TMTS %MDPI: Please use commas to separate thousands for numbers with five or more digits (not four digits) in the picture, e.g., "10000" should be "10,000".
 light curve of SDSS J013855.86+242939.2; (\textbf{b}) Lomb--Scargle periodogram of SDSS J013855.86+242939.2; (\textbf{c}) phase-folded light curve with a period of 179.0~min. 
    %with $P_1 = 179.0 \pm 2.3$ min. 
    The red line represents the best-fit model of fourth-order Fourier series.}
    \label{fig:01385585+2429393}
\end{figure}

During the TMTS observations, this source is at its high-luminosity state, with~the TMTS $L$-band magnitude being $\sim$15.4 mag. 
As shown in Figure~\ref{fig:01385585+2429393}, some flickering can be seen in its phase-folded light curve. 
A new photometric period, \mbox{$P_1 = 179.0 \pm 2.3$ min,} was revealed via the LSP.
If the period is its orbital period, J0138 is a CV located at the period-gap boundary.
However, the~{\it TESS} observations of this object (TIC~353851691) did not detect any significant periodic signal above ${\rm 2~day^{-1}}$, while the ZTF {\it r}-band and {\it g}-band LSP only present some daily aliases.
%{We queried the \textit{TESS} observations of this source (TIC 353851691) in Sector 17, but did not detect any signals except low-frequency end (i.e., $f<2 ~d^{-1}$). For ZTF r-band light curve, only daily aliases are seen.
Hence, the~physical origin of this photometric period needs further observations and~analysis.

\vspace{-12pt}

\unskip

\subsubsection{TMTS~J03471387+1611083}

\begin{figure}[H]
  \begin{adjustwidth}{-\extralength}{0cm}
  \centering
	\includegraphics[width=15.5 cm]{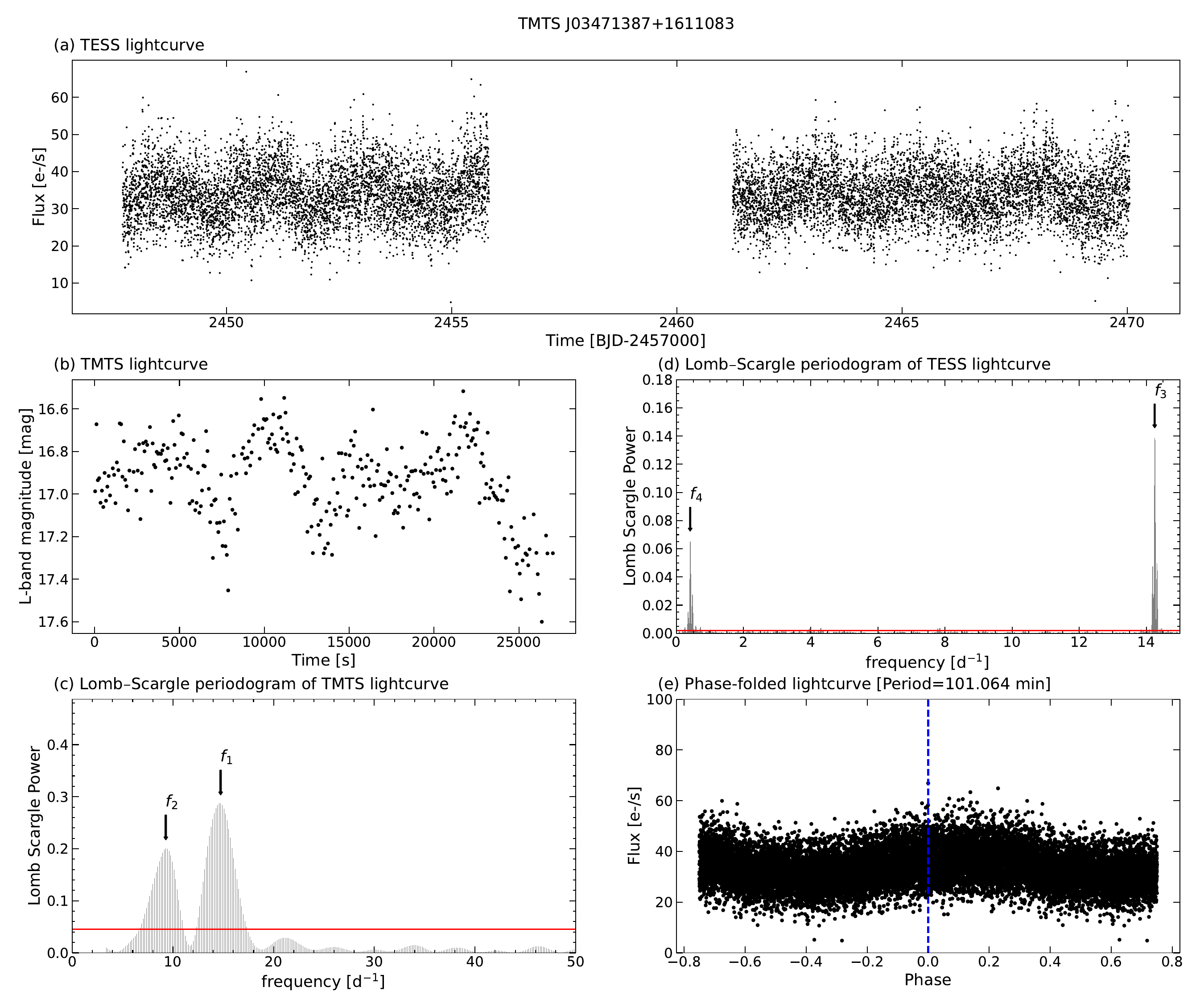}
 \end{adjustwidth}
    \caption{{(\textbf{a}) \textit{TESS} light curve of TMTS J03471387+1611083; (\textbf{b}) TMTS %MDPI: Please use commas to separate thousands for numbers with five or more digits (not four digits) in the picture, e.g., "10000" should be "10,000".
 light curve of TMTS J03471387+1611083; (\textbf{c}) Lomb--Scargle periodogram of the TMTS light curve in (\textbf{b}); (\textbf{d}) Lomb--Scargle periodogram of the \textit{TESS} light curve in (\textbf{a}); (\textbf{e}) phase-folded light curve with $P_3= 101.064$ min.}}
    \label{fig:03471387+1611083}
\end{figure}

TMTS~J03471387+1611083 (also LAMOST~J034713.84+161108.2, hereafter J0347) was first classified as an RR Lyrae star in the Catalina Survey \citep{Drake+eatl+2013+MW_100kpc}, and~then it was identified as a CV candidate with LAMOST spectroscopic observations \citep{2023AJ....165..148H}. %0.578816 day
 Following the instructions \mbox{from \citep{2023MNRAS.523.2172L},}
the Gaia absolute magnitude $M_{\rm abs, G}=6.22$~mag and dereddened \mbox{$G_{\rm Bp-Rp}=0.11$~mag} support that J0347 is a CV, rather than an RR Lyrae variable.
The TMTS light curve reveals a periodically occurring hump feature, which could be caused by a bright spot on the accretion disk.
The TMTS LSP  (bottom-left panel of Figure~\ref{fig:03471387+1611083}) presents two periodic signals, namely $P_1 = 97.7\pm 0.8~{\rm min}$ and $P_2 = 155.0 \pm 2.1~{\rm min}$. 
For comparison, we also checked the light curve of J0347 from {\it TESS} observations and computed the corresponding LSP (see the left panels of Figure~\ref{fig:03471387+1611083}).  
Two periodic signals, \mbox{$P_3 = 101.064 \pm 0.004$ min} and $P_4 =2.444 \pm 0.003$ days, were also revealed from the {\it TESS} LSP (middle-right panel of Figure~\ref{fig:03471387+1611083}). Without~an accurate determination of the orbital period, the~origin of $P_1$ and $P_3$ could not be determined reliably; a further study of J0347 is expected. 
%We inferred that TMTS period $P_1$ and TESS period $P_3$ originate from same periodic modulations of J0347, the difference could be caused by poor frequency resolution of one-night observation coverage and underestimated systematic errors of signal periods.

%However, all photometric periods obtained from both TMTS and TESS observations are inconsistent with the photometric period $P=0.578816$~day provided by Catalina Survey.

%and only found a period $P_3 = 101.064 \pm 0.004$ min, which is close to the $P_1$ measured by TMTS.

%The middle-left and bottom-left panels show the phase-folded light curves for $P_1$ and $P_2$, respectively. 
 %A signal at 2.44 d is also detected as $f_4=0.409 ~{\rm d^{-1}}$, which is the beat between $P_1$ and $P_3$ ($1/(f_3+f_4) = 98.24$ min). We suggest one of the two periods is orbit period, another one is superhump period. Without an accurate determination of the orbital period, either $P_1$ or $P_3$ could be a superhump period. Taking $P_3$ as the orbital period, the period excess $\epsilon=(P_{\rm sh}-P_{\rm orb})/P_{\rm orb}$ is $-0.033$, while assuming $P_1$ as the orbital period, the excess is 0.034. 
 %So we suggest $P_3$ is orbital period, while $P_1$ is negative superhump period, as the excess $\epsilon=-0.033$. The origin of $P_2$ is not clear, but the separation of the two humps in the light curve is about 10000 s, so we suspect $P_2$ is the modulation caused by the bright spot. 

\unskip

%%Qichun, it is not clear how 4.13-hr period was derived? from photometry?
%No, they derived two periods from radial velocity.
\subsubsection{FO~Per}
\label{FO Per}

TMTS J04083502+5114484 (FO~Per) was first discovered by \citet{1939AN....268..273M}, and~then classified as a DN with spectroscopic observations \citep{1989A&AS...78..145B}. \citet{2007PASP..119..494S} determined its orbital period as $211.2$ min or $247.8$ min by measuring the radial velocities of {H$\alpha$} emission. 
\citet{2007PASP..119..494S} suggested a 4.13~hr orbital period since the period is consistent with those of nova-like variables. However, this estimate is not conclusive, as~the orbital periods of many DNe also fall into this~range. 

\begin{figure}[H]
\begin{adjustwidth}{-\extralength}{0cm}
\centering
	\includegraphics[width=15.5 cm]{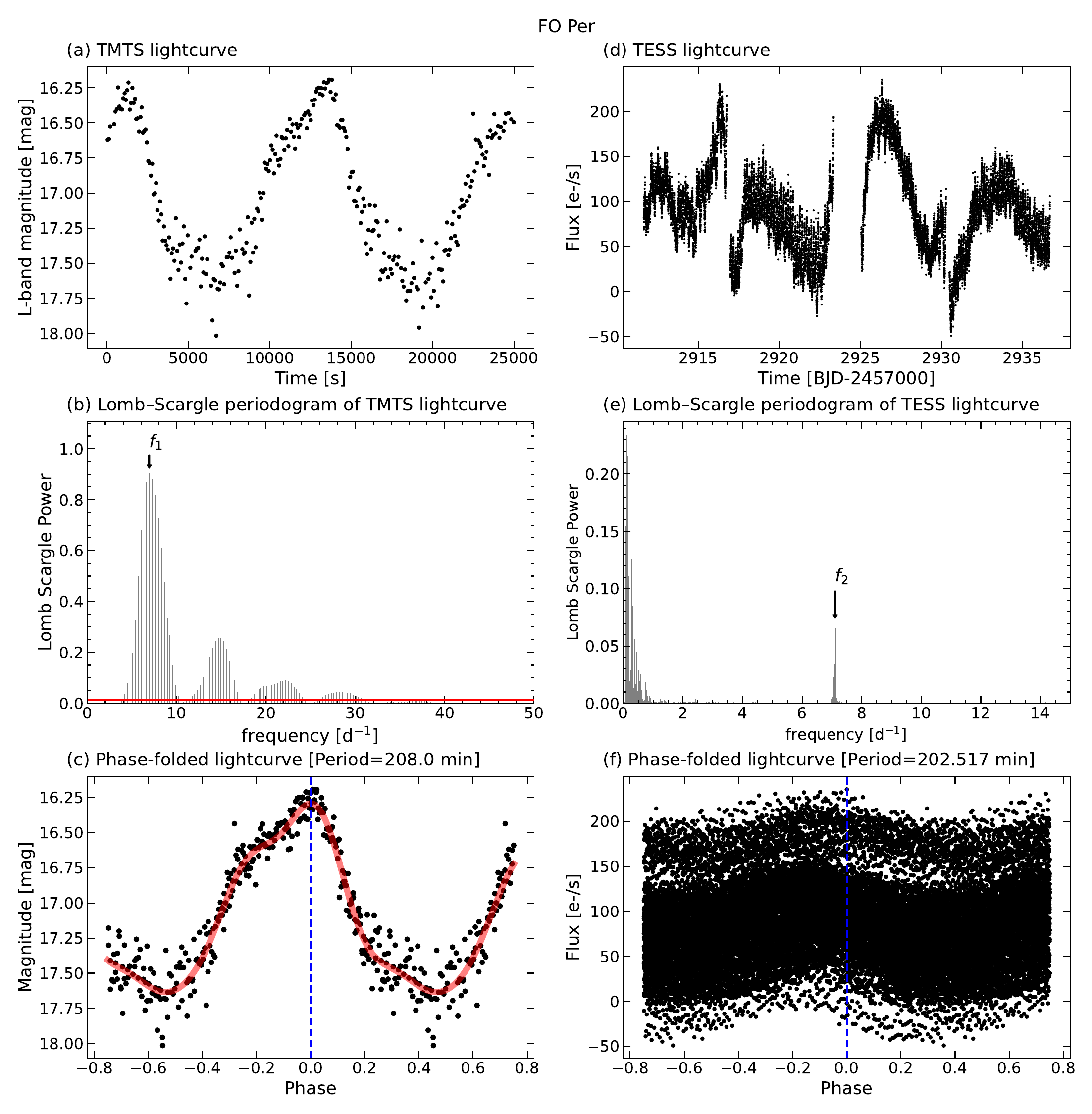}
 \end{adjustwidth}
    \caption{(\textbf{a}) TMTS light curve of FO Per; (\textbf{b}) Lomb--Scargle periodogram of the TMTS light curve; \mbox{(\textbf{c}) phase-folded} TMTS light curve with $P_1=208.0$ min, with~the red line representing the best-fit model of fourth-order Fourier series; (\textbf{d}) \textit{TESS} light curve of FO Per; (\textbf{e}) Lomb--Scargle periodogram of the \textit{TESS} light curve in (\textbf{d}); (\textbf{f}) phase-folded \textit{TESS} light curve with $P_2 = 202.517$~min. %MDPI: Please use commas to separate thousands for numbers with five or more digits (not four digits) in the picture, e.g., "10000" should be "10,000".
}
    \label{fig:04083502+5114484}
\end{figure}

In the periodograms shown in Figure~\ref{fig:04083502+5114484}, a~photometric period of $P_1=208.0 \pm 1.4$ min is detected from the TMTS observation, while another period $P_2 = 202.517 \pm 0.002$ min is detected from the {\it TESS} observation.
Both photometric periods roughly agree with the presence of a shorter orbital period suggested by \citet{2007PASP..119..494S}, implying that the orbital period of FO~Per is more likely $211.2$ min.

 %$P_1$, the period with strongest power, is consistent with 211.2 min within $3\sigma$. Thus, we suggest that $P_1$ is the orbital period while $P_2$ is a negative superhump period. This suggestion can be supported by the resulting excess $\epsilon= -0.041$ (the orbital period is taken as 211.2~min) of $P_2$.  
%is also a credible value. 
From the ZTF {\it r}-band light curve, the~brightness of TMTS J04083502+5114484 can vary from 13.5 mag to 17 mag. 
The ZTF light curve also indicates that FO~Per was going into outburst during the TMTS observation, and~this CV system then reached \mbox{$\sim$13.5~mag} ten days later. It is worth noting that the light-variation amplitude of this DN can reach  \mbox{$\sim$1.5 mag}, much larger than that of other DNe in our samples, and thus, it is possibly caused by a different physical process. 
%and should be related to the pre-outburst state.

\unskip

%%Qichun, it is not clear why you suggest P2 as a negative superhump？need some explanation
%because the period excess epsilon follows the relation between orbital period and negative superhump
\subsubsection{SS~Aur}
\begin{figure}[H]
 \begin{adjustwidth}{-\extralength}{0cm}
\centering
	\includegraphics[width=15.5 cm]{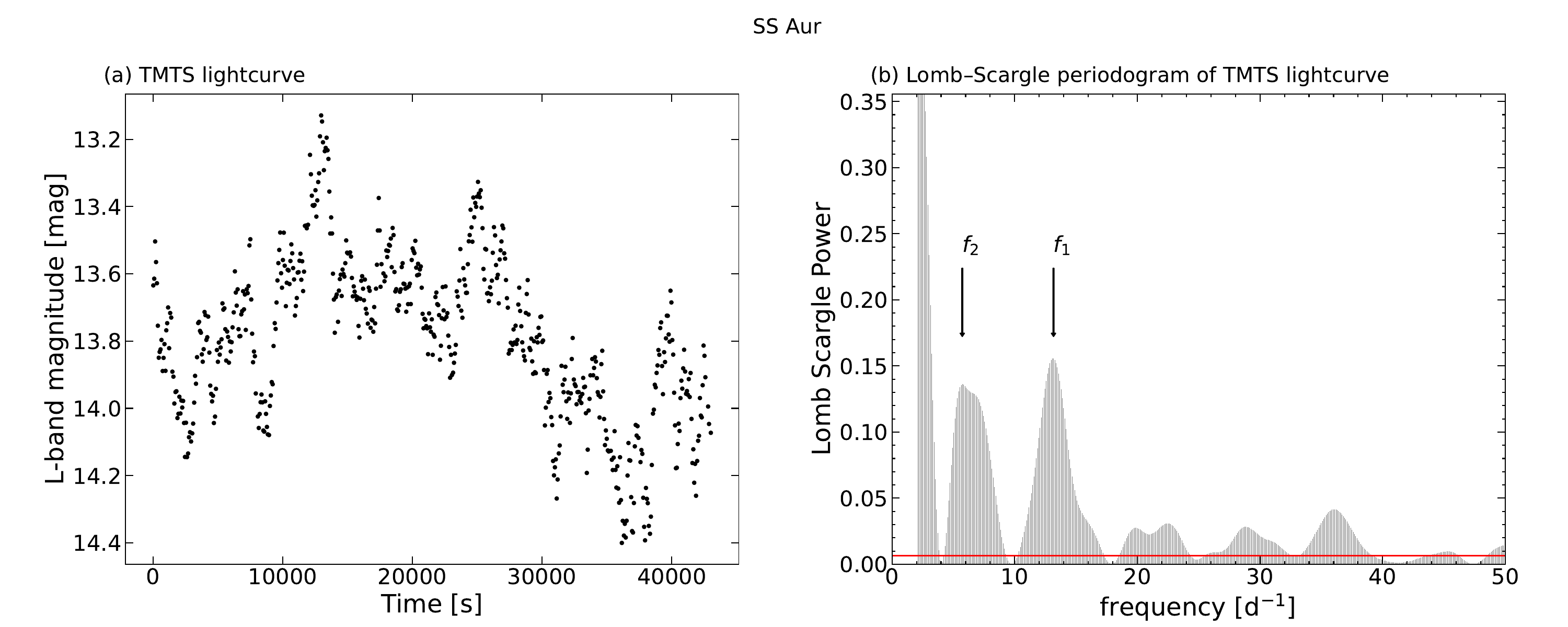}
 \end{adjustwidth}
    \caption{(\textbf{a}) TMTS %MDPI: Please use commas to separate thousands for numbers with five or more digits (not four digits) in the picture, e.g., "10000" should be "10,000".
 light curve of SS Aur; (\textbf{b}) {Lomb--Scargle}  periodogram of the TMTS light~curve. }
    \label{fig:06132238+4744248}
\end{figure}

TMTS J06132238+4744248 (SS~Aur) is a DN with an orbital period of $262.23\pm 0.15$~min derived from radial velocities \citep{1965ApJ...142.1041K,1986AJ.....92..658S, 2021ApJ...908..173G}. Two periods can be resolved in the periodogram of its TMTS light curve, namely $P_1=108.535\pm 0.015$ min and $P_2=255.83\pm 0.08$ min (see Figure~\ref{fig:06132238+4744248}). Among~these two periods,  $P_2$ is likely a negative superhump. If~so, the~period excess is $\epsilon=-0.024$, consistent with the relation between period excess and negative superhump period given by  \citet{2023MNRAS.519..352B}.
%Using the {\it TESS data}, we detect a period $P_3=1.4784\pm 0.0002$ day, which is likely the period of  three humps seen in the {\it TESS} light curve (see Figure~\ref{fig:06132238+4744248}(a)). 
\vspace{-6pt}

%Qichun, you need to list the exact date on which the polarization observation was taken below
%Dear Prof., the date is 2022 Nov.22

%%Qichun, is there explaination for DQ somewhere in the text?

\subsubsection{V378~Peg}
\begin{figure}[H]
  \begin{adjustwidth}{-\extralength}{0cm}
\centering
	\includegraphics[width=15.5 cm]{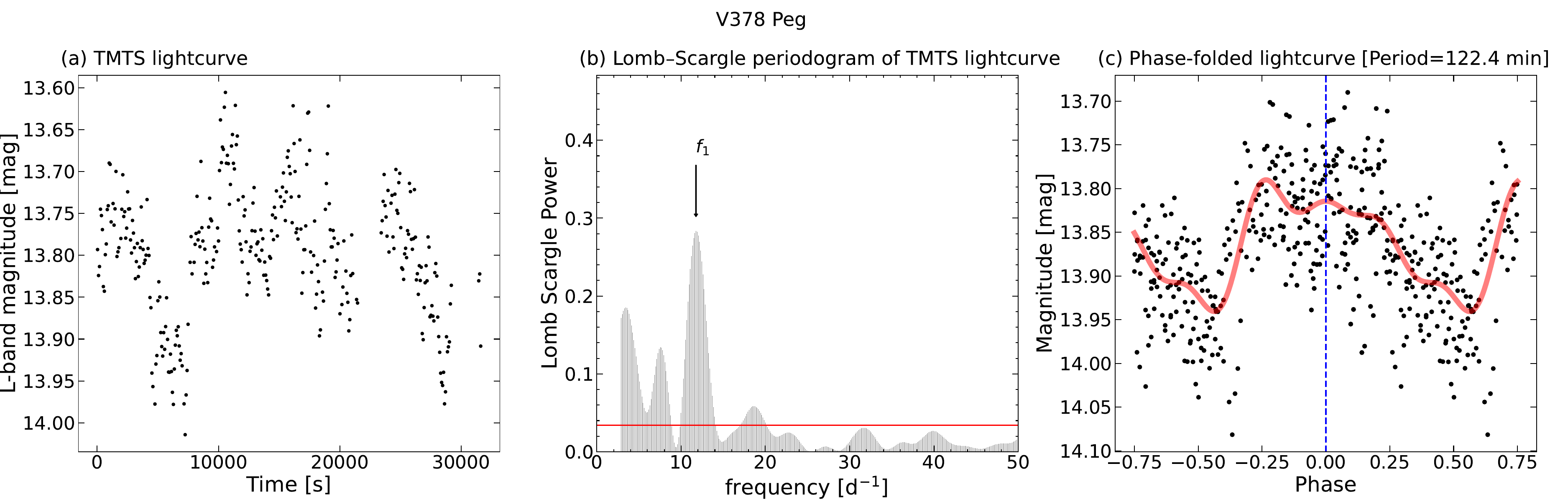}
 \end{adjustwidth}
    \caption{(\textbf{a}) TMTS light curve of V378 Peg; (\textbf{b}) Lomb--Scargle periodogram of the TMTS light curve; (\textbf{c}) phase-folded TMTS light curve with $P_1=122.4$ min (a second-order polynomial was subtracted to detrend), with~the red line representing the best-fit model of fourth-order Fourier series.  %MDPI: Please use commas to separate thousands for numbers with five or more digits (not four digits) in the picture, e.g., "10000" should be "10,000".
}
    \label{fig:23400423+3017476}
\end{figure}

TMTS J23400423+3017476 (V378 Peg) is an NL that was discovered by the Palomar--Green survey \citep{1986ApJS...61..305G} and first classified as a CV by \citet{IBVS}. The~orbital period of V378 Peg derived from radial velocities is $P_{\rm orb}=199.55\pm 0.06$ min \citep{2012NewA...17..433R}. Negative superhumps ($\sim$3.2 h) of V378 Peg were detected by both \citet{2012NewA...17..433R} and \mbox{\citet{2012NewA...17...38K}.} From~our TMTS light curve, we detected a period of $P_1=122.4 \pm 0.2$ min, which is shown in Figure~\ref{fig:23400423+3017476}. Following \citet{2018ApJS..236...16V}, we calculated the LSP of a window function of the TMTS light curve, but we did not find any fake spikes at the location of $P_1$. This period is unlikely to be a superhump, owing to the large difference with $P_{\rm orb}$. Although~we do not exactly know its physical origin, it reflects the complexity of disk oscillations in CVs. 
%Using the {\it TESS data}, we detect a period $P_3=1.4784\pm 0.0002$ day, which is likely the period of  three humps seen in the {\it TESS} light curve (see Figure~\ref{fig:06132238+4744248}(a)). 

\subsubsection{Possible~QPOs}

DNe sometimes exhibit quasi-periodic fluctuations with a period ranging from a few tens to a few thousands of seconds \citep{2021MNRAS.500.1547S}. 
Among them, the~rapid oscillations, with~a timescale of a few dozen seconds, are called dwarf nova oscillations (DNOs), while the oscillations with a longer period ranging from a few hundred to a few thousand seconds are regarded as quasi-periodic oscillations (QPOs). 
Owing to timing resolution, TMTS cannot detect any DNOs; instead, some possible QPOs were detected from the uninterrupted light curves of TMTS.
%such as TMTS J01043552+4117576, which are shown in Figure~\ref{fig: J01043552+4117576}. 

\begin{figure}[H]
 \begin{adjustwidth}{-\extralength}{0cm}
\centering
	\includegraphics[width=15.5 cm]{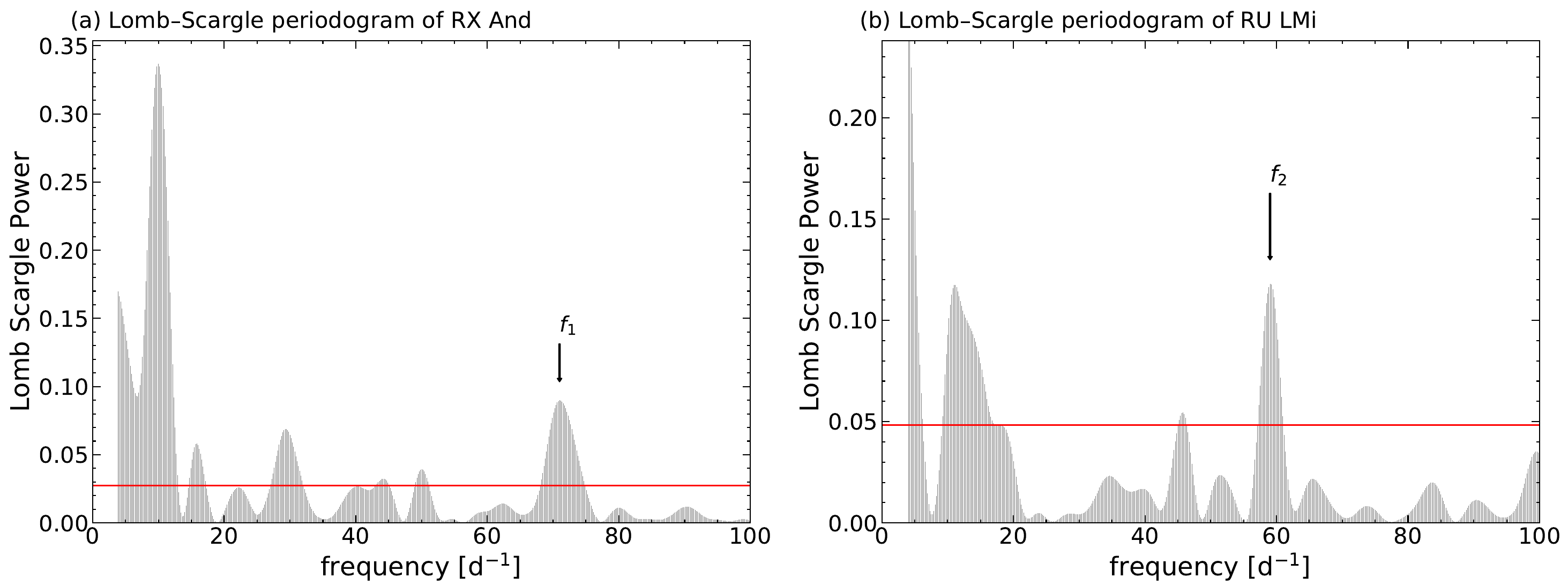}
 \end{adjustwidth}
    \caption{(\textbf{a}) Lomb--Scargle periodogram of RX And; (\textbf{b}) Lomb--Scargle periodogram of RU LMi. %MDPI: please confirm if red line need further explanation.
}
    \label{fig: J01043552+4117576}
\end{figure}

TMTS J01043552+4117576 (RX And) is a Z~Cam-type DN with an orbital period of 5.08~h \citep{1962ApJ...135..408K}. A~periodicity of 35.7~s (DNO) was found in previous light curves \citep{1976ApJ...207..190S}, while \mbox{\citet{2003MNRAS.344.1193W}} proposed that there was weak evidence of a 1000~s oscillation. In~the upper panel of Figure~\ref{fig: J01043552+4117576}, the~LSP of the TMTS observations presents a period signal $P_1=20.285\pm 0.006$ min, within~the typical timescale of~QPOs.

\textls[-12]{TMTS J10020745+3351005 (RU LMi) was initially identified as a CV by \citet{1988ApJ...328..213W}.} With~photometric observations, \citet{1990PASP..102..758H} reported an orbital period of 355 min, while the spectrum of this object was consistent with those of typical DNs \citep{1994IBVS.4074....1H}. 
In the lower panel of Figure~\ref{fig: J01043552+4117576}, a~period signal of $P_2=24.40\pm 0.07$~min was detected via TMTS, which is also within the typical timescale of~QPOs.

%QPOs of polars also attracted many interests, especially the oscillations with frequency of a Hz order, due to the thermal instability in the post-shock accretion column~\citep{2015A&A...579A..24B, 2018MNRAS.474.1629B}. 

%%Qichun，it is not complete to say "Reflection effect can explain some high-amplitude light variations seen in the light curves", you need to mention "in the light curve of what ..."; you mention "a bright spot" below, any references for such a statement?
%Dear Prof., I add "the light curves of periodic variable stars";
\section{Discussion}
\unskip

\subsection{Statistical Properties of {\rm~EW} and {\rm FWHM} for Nonmagnetic CVs}

%\subsection{Equivalent Width and FWHM}
\label{ew_Halpha}
As introduced in Section~\ref{spec}, we collected the spectra of 44 CVs, for~which the EW and FWHM of H${\alpha}$ emission were measured. Since the accretion in magnetic CVs is governed by magnetic fields, their accretion geometries are essentially different from those of nonmagnetic systems in which the radiation of accretion disks dominates the emission \citep{1990SSRv...54..195C, 2020MNRAS.495.4445K}. 
Here, we focus on the spectroscopic properties of the 28 nonmagnetic~CVs.

According to the period gap \citep{2006MNRAS.373..484K}, the~CVs can be divided into two groups: short-period CVs with $P_{\rm orb} < 2.15$~h and long-period CVs with $P_{\rm orb} > 3.18$~h. Adopting the orbital periods provided from the VSX shows that %Please check that the intended meaning has been retained.
 the~28 nonmagnetic CVs consist of 19 long-period systems (10 DNe and 9 NLs) and 9 short-period ones (9 DNe). 
%We show the distributions of ${\rm EW}_{\rm H\alpha}$ of the three subclasses respectively. 

Because the sample of nonmagnetic CVs from the TMTS observations is relatively small, spectral data from the SDSS were also included in double-checking the robustness of the distribution trend. The~SDSS CVs were cross-matched with the VSX catalog to collect their identification information, such as the subtype and orbital period. Then, the SDSS spectra of those nonmagnetic CVs (DNe and NLs) were analyzed with the same criteria as described in Section~\ref{spec}. 

The spectra of DNe in the outburst state usually show narrow Balmer emission lines superposed on broad absorption wings.
We adopted a composite model, a~positive Gaussian function for the emission feature plus a negative one for the absorption, to~fit such a profile.
%a model-- one Gaussian function for absorption and one for emission was adopted to fit $F_{\rm sub}(\lambda)$, and then the absorption component was removed as a correction, which is supposed to have a different physical origin compared with the emission component \citep{2020AJ....159...35W}. 
We found that the emission components during the outburst tend to have a small EW (typically, ${\rm EW}_{{\rm H}\alpha}<10$ \AA).
For example, 
%However, we found that the emission components all have a small ${\rm EW}_{\rm H\alpha}$ (i.e., ${\rm EW}_{\rm H\alpha}<10$ \AA). 
%What's more, 
the ${\rm EW}_{{\rm H}\alpha}$ of SDSS J105550.08+095620.4 is 8.54 \AA ~during the outburst, while its ${\rm EW}_{{\rm H}\alpha}$ was reported to be 70 \AA\ in quiescence \citep{2016AJ....152..226T}.
The significant difference in EW between the outburst and quiescent states prevented us from performing a reasonable comparison of the emission strengths of DNe.
%indicating the ${\rm H\alpha}$ emissions in quiescence and outburst differ a lot. 
We thus excluded the DNe  in the outburst state from our statistical~study.

%we convolved the DN SDSS spectra with a \textit{g}-band filter transmission curve to get the \textit{g}-band magnitudes, and compared to the long-term ASAS-SN \textit{g}-band light curves \citep{2014ApJ...788...48S, 2023arXiv230403791H}. 
%The sources in the outburst state are not taken into our statistic study.

\begin{figure}[H]
	\includegraphics[width=10.5 cm]{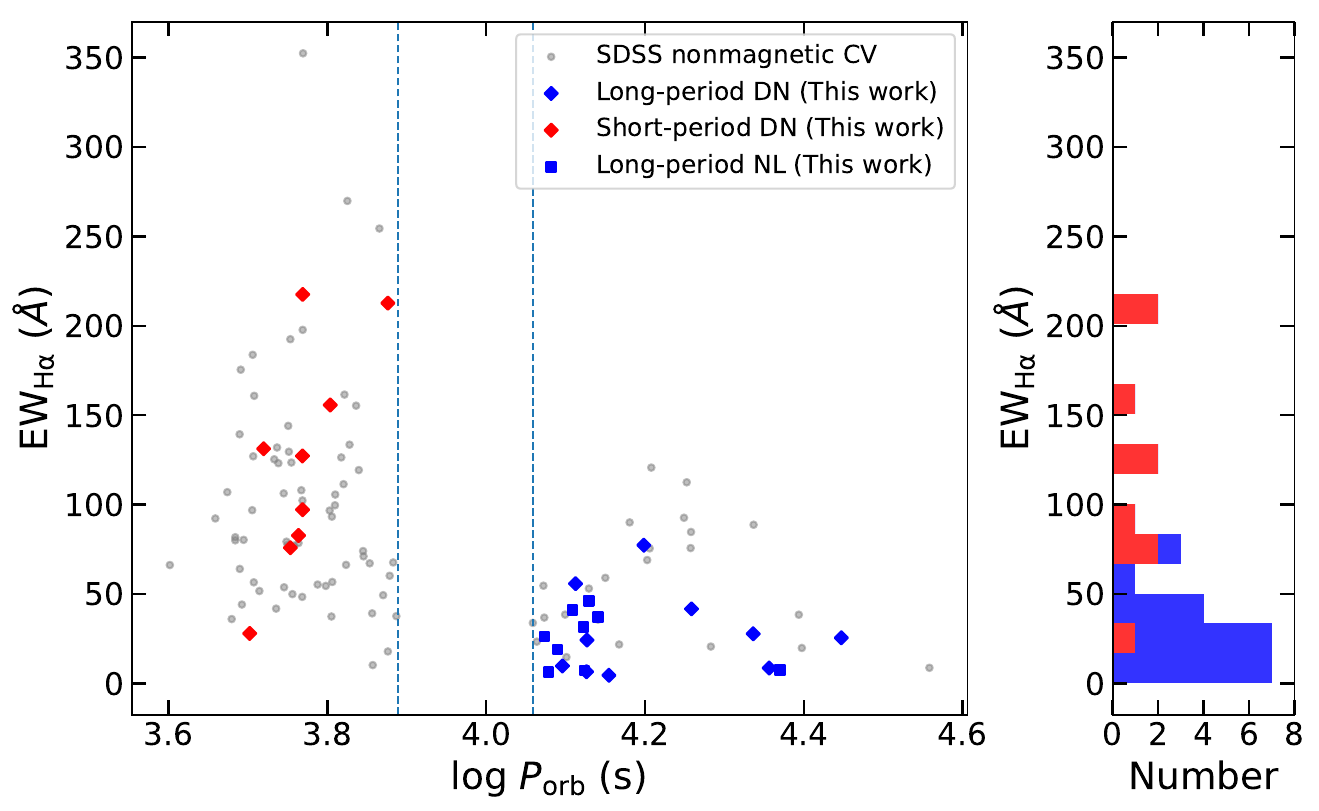}
    \caption{Distribution of ${\rm EW}_{{\rm H}\alpha}$ measured for the nonmagnetic CV sample as a function of $\log{P_{\rm orb}}$. Diamond and square points represent the DNe and NLs in this catalog, respectively, while red and blue colors indicate short-period and long-period subclasses. Gray circle points represent the sources from \citet{2011AJ....142..181S}. The~two vertical blue dashed lines mark the period gap \mbox{($2.15~{\rm h} \lesssim P_{\rm orb} \lesssim 3.18~{\rm h}$}; see \citet{2006MNRAS.373..484K}).}
    \label{fig:EW_period}
\end{figure}

\begin{figure}[H]
 \begin{adjustwidth}{-\extralength}{0cm}
\centering
	\includegraphics[width=15.5 cm]{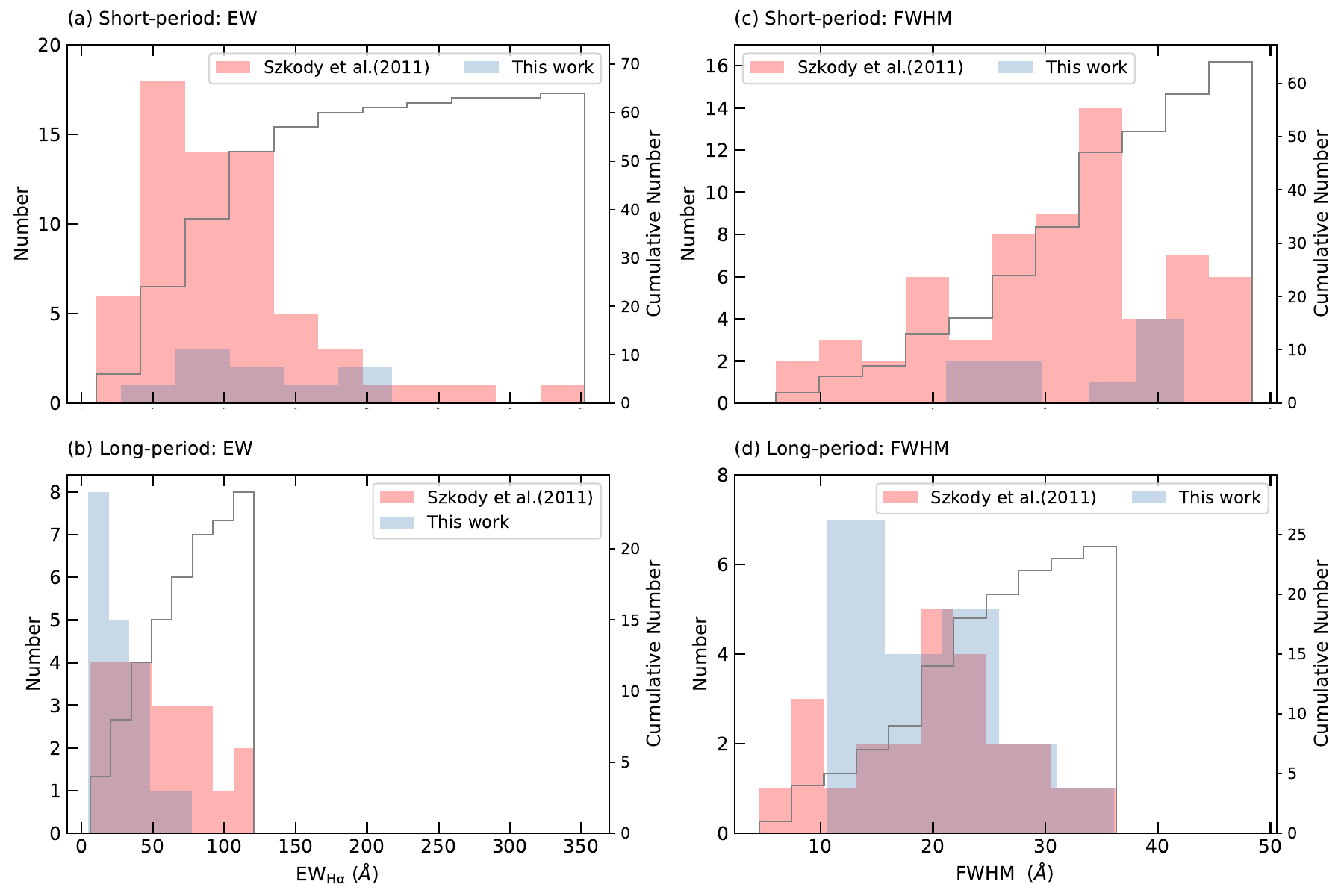}
 \end{adjustwidth}
    \caption{Histogram distribution of EW (\textbf{a},\textbf{b}) and FWHM (\textbf{c},\textbf{d}) of H${\alpha}$ emission in spectra of nonmagnetic CVs. Panels (\textbf{a},\textbf{c}): distribution of short-period nonmagnetic CVs. Panels (\textbf{b},\textbf{d}): distribution of long-period nonmagnetic CVs. The~red columns represent the nonmagnetic CVs from \mbox{\citet{2011AJ....142..181S}}, while the blue columns represent the nonmagnetic CVs from our sample. The~gray lines show the cumulative distributions of the SDSS nonmagnetic CV~sample.}
    
   % (a): blue line represents the histogram of ${\rm EW}_{\rm H\alpha}$ for short-period CVs, red line represents the corresponding fitted distribution. (b): blue line represents the histogram of ${\rm EW}_{\rm H\alpha}$ for gap-period CVs, red line represents the corresponding fitted distribution. (c): blue line represents the histogram of ${\rm EW}_{\rm H\alpha}$ for long-period CVs, red line represents the corresponding fitted distribution. (d): blue line represents the histogram of ${\rm FWHM}$ for short-period CVs, red line represents the corresponding fitted distribution. (e): blue line represents the histogram of ${\rm FWHM}$ for gap-period CVs, red line represents the corresponding fitted distribution. (f): blue line represents the histogram of ${\rm FWHM}$ for long-period CVs, red line represents the corresponding fitted distribution.}
    \label{fig:loggauss}
\end{figure}

For the TMTS sample, the~distributions of ${\rm EW}_{{\rm H}\alpha}$ are shown in Figure~\ref{fig:EW_period}, in~which the nonmagnetic CVs with shorter orbital periods tend to have a wider feasible range of ${\rm EW}_{{\rm H}\alpha}$ than those with longer periods.
%stronger ${\rm H\alpha}$ emission tend to have shorter orbital periods. 
The mean ${\rm EW}_{{\rm H}\alpha}$ values of CVs in the two subclasses are 125 ~\AA ~and 26 ~\AA, respectively. 
A similar trend can be seen from the ${\rm FWHM}_{{\rm H}\alpha}$ distributions (see the blue columns in Figure~\ref{fig:loggauss}).

\vspace{-6pt}

\vspace{-12pt}

%Considering the size of our non-magnetic CV sample is limited, the SDSS CV sample \citep{2011AJ....142..181S} were adopted to verify whether these differences are real. 

The EW/FWHM distributions for 64 short-period CVs and 24 long-period CVs obtained from SDSS observations are appended to Figure~\ref{fig:loggauss} (the red columns; see also Table~\ref{tab2}). 
%The final SDSS non-magnetic CV sample contains 64 short-period sources (63 DNe, 1 NL) and 24 long-period sources (14 DNe, 10 NLs). The distribution of ${\rm EW}_{\rm H\alpha}$ against orbital period is shown in Figure~\ref{fig:EW_period}.
%The observed distributions are shown in Figure~\ref{fig:loggauss} in red, and 
%Table.\ref{tab2} presents the statistical parameters  for the distributions. 
The distribution trend yielded from the SDSS spectra is consistent with that revealed in the TMTS CV samples. 
%The high mean and standard variance of ${\rm EW}_{\rm H\alpha}$ found in short-period non-magnetic CVs are consistent with the finding in our sample, confirming the short-period non-magnetic CVs and the long-period non-magnetic  CVs show significant discrepancy in the strength of H${\alpha}$ emission. 
To quantify the difference in H${\alpha}$ strength between the long-period and short-period nonmagnetic CVs, 
%the SDSS ${\rm EW}_{\rm H\alpha}$ distributions, 
we performed the two-sample Kolmogorov--Smirnov test (K-S test) for the SDSS ${\rm EW}_{{\rm H}\alpha}$ distributions and obtained a $p$-value of 0.003, suggesting that the ${\rm EW}_{{\rm H}\alpha}$ distributions above/below the period gap are quite different.
Similarly, the~$p$-value obtained for the  SDSS ${\rm FWHM}_{{\rm H}\alpha}$ distribution is~0.00005.

\begin{table}[H]
\caption{\textls[-15]{Summary of the statistics of the distributions for the short-period and long-period nonmagnetic CVs from our sample and \citet{2011AJ....142..181S}. {The subscripts ``mean'' and ``std'' respectively represent the mean values and standard deviations of EW/FWHM for short-period and long-period~classes.}} \label{tab2}}
%	\begin{adjustwidth}{-\extralength}{0cm}
		\begin{tabularx}{\textwidth}{Cp{2cm}<{\centering}ccc}
			\toprule
			\textbf{Class} &   \boldmath{${\rm EW_{\rm mean}}$}  &\boldmath{${\rm EW}_{\rm std}$} & \boldmath{${\rm FWHM_{\rm mean}}$}  &  \boldmath{${\rm FWHM}_{\rm std}$}\\
  & \textbf{(\AA)} & \textbf{(\AA)}  & \textbf{(\AA)} & \textbf{(\AA)} \\
			\midrule
Short-period (\citet{2011AJ....142..181S})  & 101.19 &  60.86  & 30.96 & 10.16 \\
  Long-period (\citet{2011AJ....142..181S})  & 53.07 & 32.70  & 19.90 & 8.07\\
  Short-period (this work) & 125.33 &  59.30  & 33.55 & 7.26 \\
   Long-period (this work)  & 26.47 &  19.33  & 19.47 & 6.96 \\
			\bottomrule
		\end{tabularx}
%	\end{adjustwidth}
	%\noindent{\footnotesize{* Tables may have a footer.}}
\end{table}

%(see Figure~\ref{fig:EW_period}). 
%Intuitively, this is likely related to the evolution of CV, which leads to stronger ${\rm H\alpha}$ emission intrinsic to some sources with shorter periods. 
As shown in Figure~\ref{fig:loggauss}, the~EW and FWHM of the short-period nonmagnetic CVs are systematically larger than those of long-period systems. In~particular, Figure~\ref{fig:EW_period} demonstrates that only short-period systems exhibit large ${\rm EW}_{{\rm H}\alpha}$ (e.g., $\gtrsim$130~\AA) in their spectra. 
%The larger FWHM observed for some short-period non-magnetic CVs 
%implying that their ${\rm H\alpha}$ lines undergo more intense broadening. 
We supposed that the accretion disks in short-period nonmagnetic CVs tend to have smaller radii, which means that 
%We suppose that the accretion disc of the non-magnetic CVs might have a  smaller radius when they evolve into the short-period branch. Thus 
the H$\alpha$ emission regions are potentially closer to the accretors {and thus suffer from a more intense Kepler 
broadening.} In Figure~\ref{fig:EW_period}, the~H$\alpha$ strength among long-period sources does not show a correlation with $P_{\rm orb}$, which is a clue to the possible transformation of the accretion process that occurs when CVs go through the period gap.

Although the evolutionary path of CVs has been well understood \citep{2011ApJS..194...28K}, there are still some challenges in interpreting the large spread of CV orbital period distribution and the presence of more massive WDs in CVs \citep{2017MNRAS.466.2855P, 2022MNRAS.510.6110P}. These facts suggest that our
understanding of CV evolution is incomplete. The~distributions of H$\alpha$ strength presented here shed  light on the CV evolution.
%Unlike possible long timescale trackers of a CV's evolution such as the donor radius, accretion brightness, and WD temperature \citep{2011ApJS..194...28K}, 
%the high-cadence light curves and spectral emission lines may represent short-timescale dynamic variations in CV systems. 

Since nonmagnetic CVs in the TMTS CV samples present the same ${\rm EW}_{{\rm H}\alpha}$ and ${\rm FWHM}_{{\rm H}\alpha}$ tendency as those from the SDSS sample, the~effect of selection bias on the above conclusion is unlikely to be significant. Note that the SDSS CV sample should be more homogeneous since it includes faint sources and covers a wide color range \citep{2009MNRAS.397.2170G, 2011A&A...536A..42Z}. %After cross-matching the SDSS non-magnetic CV sample with Gaia \citep{2018A&A...616A...1G} sources within an aperture of radius $3.^{\prime \prime}0$, we found that the short-period CVs tend to have darker apparent G-band magnitude in comparison to the long-period sources, with~a limiting magnitude $\sim$ 20 mag.  
%The CV sample clearly does not be bent towards bright end based on our criteria (i.e., sources with high quality spectra). 
%Up to date, the number of non-magnetic CVs having available spectroscopic data are not sufficient, so 
However, the~statistical conclusions in this work will require further verification with a more complete sample in the~future.

\subsection{The Implications of TMTS Light~Curves}
In Section~\ref{lc_feature}, we  highlighted several CV samples showing abnormally large-amplitude periodic variations. 
%From Figure~\ref{fig:lc5} we can see that the light curves of polars 
The polar EV~UMa exhibits a large-amplitude orbital modulation, which can be explained by a heavily beamed cyclotron emission from the accretion column. The~change in the angle between the line of sight and magnetic polar within an orbital period leads to large-amplitude photometric variations in the light curves \citep{2001A&A...372..557G, 2015IBVS.6129....1L}. 
Differing from most DNe in our samples, the~dwarf nova FO~Per presents a large-amplitude orbital modulation comparable to the polars. 
%In Section~\ref{FO Per}, we interpret this variation as an orbital modulation which possibly relate to the pre-outburst state. 
%Reflection effects can explain some high-amplitude periodic variations seen in the light curves of binary systems \citep{2022Natur.605...41B}, which were studied in several CVs \citep{2018AJ....156..153D}. However, we did not find any colour-dependent modulation for FO~Per from the ZTF $g$-band and $r$-band light curves. 

\begin{figure}[H]
 \begin{adjustwidth}{-\extralength}{0cm}
\centering
	\includegraphics[width=15.5 cm]{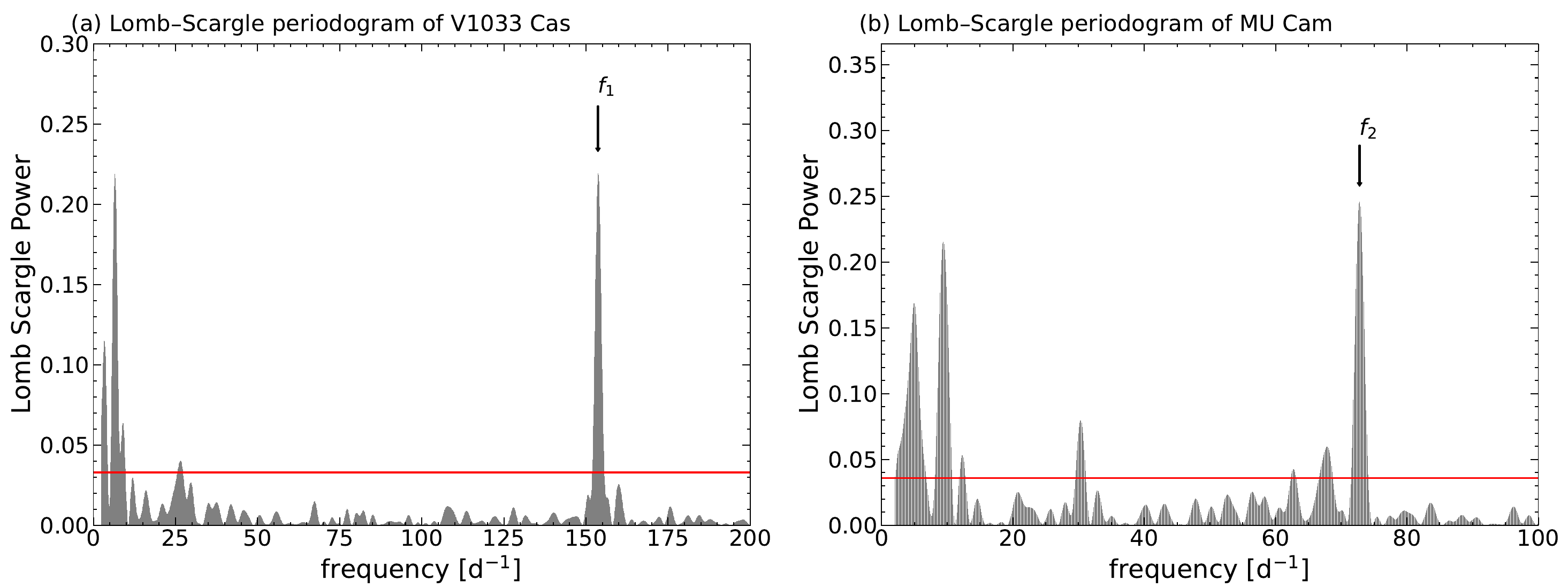}
 \end{adjustwidth}
    \caption{(\textbf{a}) Lomb--Scargle periodogram of V1033 Cas; (\textbf{b}) Lomb--Scargle periodogram of MU Cam; $f_1$ and $f_2$ indicate the spin frequencies of the two sources, respectively. %MDPI: please confirm if red line in figure need further explanation.
}
    \label{fig: spin}
\end{figure}

Since the periodic variations could be related to the specific accretion state of CVs (e.g., pre-outburst state), the~high-cadence survey observations are crucial to uninterruptedly record the light variations corresponding to the short-duration states in  CVs (e.g., state transition). Furthermore, the~high-cadence observations from TMTS can easily capture the rapid periodic variations (i.e., spin modulations) from IPs.
For example, two spin frequencies for V1033 Cas ($P_{\rm pho}=9.372\pm 0.005$ min) and MU Cam (\mbox{$P_{\rm pho}=19.785\pm 0.009$ min}) were also revealed through the TMTS data (see Figure~\ref{fig: spin}).
%in the sources with rapid periodic variations there are some IPs that TMTS light curves captured their spin-period modulations, like  $P_{\rm pho}=9.372\pm 0.005$ min for V1033 Cas (TMTS J00225764+6141076; \citet{2007A&A...473..185B, 2009A&A...501.1047A}) and $P_{\rm pho}=19.785\pm 0.009$ min for MU Cam (TMTS J06251631+7334386; \citet{2003A&A...406..213A, 2003A&A...406..253S}). The two spin frequencies are shown in Figure~\ref{fig: spin} with strong powers. 
%That means that, by searching ultra-short periodic signals, photometric surveys can discover potential IP candidates.

%We also examine other DNs with large orbital amplitudes, such as TMTS J07485955+3125121 (one of the SDSS sources; \citealt{2023MNRAS.524.4867I}) which has an amplitude of $0.45$ mag. 

%Such a light variation is mainly caused by a hump seen during an orbital cycle caused by a bright spot. A similar case applies to another CV system, TMTS J09121621+5053531, with an amplitude of $0.34$ mag. This kind of orbital hump is a common photometric behaviour of CVs \citep{1971MNRAS.152..219W, 2018AJ....155...61R}.

%One thing needs to be point out is that due to the short observation baseline, from TMTS light curves we can not identify some characteristic features like outbursts and state transitions of CVs, thus limiting us to discover new candidates by our own. 
\vspace{-6pt}
\unskip

%%Qichun, the last paragraph seems incomplete, may need a few sentences to address the future study of these peculiar CVs and their impact to understanding of CV physics
\section{Conclusions}
We {have presented} well-sampled light curves and spectra of 64 CVs or CV candidates observed/discovered during the first 3~years of the TMTS survey. %Taking advantage of the increasing number of known orbital periods, we studied the distributions of orbital amplitudes. 
\textls[-15]{By performing periodogram analysis, we {identified} two new CV candidates (TMTS J04405040+6820355 and  TMTS J06183036\\+5105550) and nine new photometric periods for seven known CVs from the TMTS light curves. The~properties of the two new CVs and the physical origins of new periods {were} discussed. 
TMTS J04405040+6820355 is inferred to be a new DN candidate, while TMTS J06183036+5105550} could be a new IP, as indicated by a linear polarization of $\sim$0.6\%.

%As introduced in Section~\ref{individual}, due to the absence of  He \Rmnum{2} $\lambda$4686 line, TMTS J04405040+6820355 is inferred to be a new DN candidate; TMTS J06183036+5105550, which presents a linear polarization of $\sim 0.6\%$, could be a new IP.

The short-timescale features from our high-cadence light curves can be  classified into four types: eclipse, low-amplitude periodic variation, high-amplitude periodic variation, and~rapid periodic variation. We {attempted to explore the various short-timescale variations in CVs with these features.}
%apply these features to explore the fast-developing physical process in CVs, which help the CV subclassification.

%we revealed several periodic photometric variations for CV sources, and classified them into four classes. These sources may work as test targets for physical models of SHs, QPOs, and so on in the future, and some of them are worth further detailed study. As one of the sources with largest amplitude in our sample, we highlighted FO~Per and contributed this variation to orbital modulation prior to outburst state. 
%We investigated the new CV candidates TMTS J04405040+6820355 and TMTS J06183036+5105550, the last of which is a IP system containing a magnetic WD, showing a linear polarization of $\sim 0.6\%$. The rapid periodic variations detected by TMTS imply that high-cadence photometric surveys can discover short signals like QPO and rotation modulations from IPs.

%We also reported some negative superhumps, {oscillations from unknown origin}, and possible QPOs found in the light curves.  

%with TMTS high-cadence observations, revealing notable photometric variations for some sources. 

With the CV spectra from LAMOST, XLT, and~SDSS, we found that there are significant differencs in the distribution of H${\alpha}$ emissions (i.e., ${\rm EW}_{{\rm H}\alpha}$ and ${\rm FWHM}_{{\rm H}\alpha}$) between nonmagnetic CVs located above and below the period gap, implying that the accretion nature of CVs should be tightly related to their evolutionary~stages.

%These observational facts help us understand the accretion disc in non-magnetic CV systems, providing essential clues to the origin of ${\rm H\alpha}$ emission and possible transformation of accretion process occurring from long-period CVs to short-period CVs. 

%The distribution difference implies that, the short-period CVs may have smaller radii of their discs and the ${\rm H\a}$

%short-period non-magnetic CVs and long-period non-magnetic CVs, with the former having a larger median/mean value and variation, 

%we explored the H${\alpha}$ emission (i.e., ${\rm EW}_{\rm H\alpha}$) for our CV sample and SDSS CV sample. 

%Along with the SDSS CV sample, we found that the distributions of ${\rm EW}_{\rm H\alpha}$ and ${\rm FWHM}$ in short-period non-magnetic CVs are quite different from those in long-period non-magnetic CVs, with the former having a larger median/mean value and variation,
%CVs with large ${\rm EW}_{\rm H\alpha}$ (i.e., $> 130$~\AA) all have periods below the gap, 
%likely owing to the smaller radii of their accretion discs. 

%open a window for the origin of ${\rm H\alpha}$ emission and detailed transformation of accretion process occurring from long-period CVs to short-period CVs.
\vspace{6pt}

\authorcontributions{Conceptualization and methodology, Q.L., J.L. (Jie Lin), and X.W.; data analysis, Q.L.; resources, Y.S., G.X., J.M., J.L. (Jialian Liu), S.Y., A.V.F., T.G.B., Y.Y., K.C.P., Y.C., Z.C., L.C., F.G., X.J., G.L., W.L. (Wenxiong Li), W.L. (Weili Lin), C.M., X.M., H.P., Q.X., D.X., and J.Z.; writing---original draft preparation, Q.L.; writing---review and editing, J.L. (Jie Lin), X.W., Z.D., and A.V.F. All authors have read and agreed to the published version of the~manuscript.}

\funding{This %MDPI: Information regarding the funder and the funding number should be provided. Please check the accuracy of funding data and any other information carefully.
 work was supported by the National Natural Science Foundation of China (NSFC grants 12288102 and 12033003), the~Ma Huateng Foundation, and~the New Cornerstone Science Foundation through the XPLORER PRIZE. J.L. is supported by the Cyrus Chun Ying Tang Foundations. A.V.F.'s team at U.C. Berkeley received support from the Christopher R. Redlich Fund, Gary and Cynthia Bengier, Clark and Sharon Winslow, Sanford Robertson, Alan Eustace, Briggs and Kathleen Wood, and~many other donors. 
% Y.Y. appreciates generous financial support provided to the supernova group at U.C. Berkeley by Gary and Cynthia Bengier, Clark and Sharon Winslow, Sanford Robertson, and numerous other donors. 
Y.-Z. Cai is supported by 
% the National Natural Science Foundation of China 
NSFC grant 12303054 and the International Centre of Supernovae, Yunnan Key Laboratory (No. 202302AN360001).}

\dataavailability{The study's catalogs are all available in this paper. %Please check that the intended meaning has been retained.
The TMTS photometric data, Xinglong 2.16 m telescope spectra, and~Lick 3 m Shane telescope spectra can be obtained by contacting the corresponding authors.}  %MDPI: Please check if the information regarding data in the Acknowledgments Section should be moved here. If yes, please revise. For guidelines of the Data Availability Statement, please visit https://www.mdpi.com/ethics#_bookmark21.

% Only for journal Nursing Reports
%\publicinvolvement{Please describe how the public (patients, consumers, carers) were involved in the research. Consider reporting against the GRIPP2 (Guidance for Reporting Involvement of Patients and the Public) checklist. If the public were not involved in any aspect of the research add: ``No public involvement in any aspect of this research''.}

% Only for journal Nursing Reports
%\guidelinesstandards{Please add a statement indicating which reporting guideline was used when drafting the report. For example, ``This manuscript was drafted against the XXX (the full name of reporting guidelines and citation) for XXX (type of research) research''. A complete list of reporting guidelines can be accessed via the equator network: \url{https://www.equator-network.org/}.}

% Only for journal Nursing Reports
%\useofartificialintelligence{Please describe in detail any and all uses of artificial intelligence (AI) or AI-assisted tools used in the preparation of the manuscript. This may include, but is not limited to, language translation, language editing and grammar, or generating text. Alternatively, please state that “AI or AI-assisted tools were not used in drafting any aspect of this manuscript”.}
\newpage
\acknowledgments{This work includes the data from LAMOST (the Large Sky Area Multi-Object Fiber Spectroscopic Telescope), which is a National Major Scientific Project built by the Chinese Academy of Sciences. Funding for the project was provided by the National Development and Reform Commission. This work made use of data from the publicly available SDSS  %MDPI: Please check if the information regarding data in this section should be moved to Data Availability Statement section, and information regarding funding moved to Funding section. If yes, please revise.
12 data release. Funding for the Sloan Digital Sky Survey IV was provided by the Alfred P. Sloan Foundation, the~U.S. Department of Energy Office of Science, and~participating institutions. SDSS~IV acknowledges support and resources from the Center for High Performance Computing at the University of Utah. The~SDSS website is %MDPI: Please provide the access date of the URL in the following format: "URL (accessed on Day Month Year)".
 \url{www.sdss4.org}. This paper includes data collected through the {\it~TESS} mission, obtained from the MAST data archive at the Space Telescope Science
Institute (STScI). Funding for the {\it TESS} mission is provided by the
NASA Explorer Program. STScI is operated by the Association of
Universities for Research in Astronomy, Inc., under~NASA contract
NAS 5--26555. We also used data from the European Space Agency (ESA) mission Gaia %MDPI: Please provide the access date of the URL in the following format: "URL (accessed on Day Month Year)".
 (\url{https://www.cosmos.esa.int/gaia}), processed by the Gaia Data Processing and Analysis Consortium (DPAC, %MDPI: Please provide the access date of the URL in the following format: "URL (accessed on Day Month Year)".
 \url{https://www.cosmos.esa.int/web/gaia/ dpac/consortium}). Funding for the DPAC was provided by national institutions,
in particular the institutions participating in the Gaia Multilateral Agreement. 
This work made use of observations obtained with the Samuel Oschin 48-inch telescope and the 60-inch telescope at Palomar Observatory as part of the Zwicky Transient Facility project. ZTF is supported by the U.S. National Science Foundation (NSF) under grants AST-1440341 and AST-2034437 and~collaboration including current partners Caltech, IPAC, the~Weizmann Institute for Science, the~Oskar Klein Center at Stockholm University, the~University of Maryland, Deutsches Elektronen-Synchrotron and Humboldt University, the~TANGO Consortium of Taiwan, the~University of Wisconsin at Milwaukee, Trinity College Dublin, Lawrence Livermore National Laboratories, IN2P3,
University of Warwick, Ruhr University Bochum, and~Northwestern University, and~former partners the University of Washington,
Los Alamos National Laboratories, and~Lawrence Berkeley National
Laboratories. Operations are conducted by COO, IPAC, and~UW. This research made use of the International Variable Star Index
(VSX; \citet{2006SASS...25...47W}) database, operated at AAVSO, Cambridge,
MA, USA.}

\conflictsofinterest{The authors declare no conflicts of~interest.}

\appendixtitles{no} % Leave argument "no" if all appendix headings stay EMPTY (then no dot is printed after "Appendix A"). If~the appendix sections contain a heading then change the argument to "yes".
\appendixstart
\appendix
\renewcommand\thefigure{A\arabic{figure}}   

\section[\appendixname~\thesection]{}
\vspace{-6pt}
\begin{figure}[H]
 \begin{adjustwidth}{-\extralength}{0cm}
    \centering
	\includegraphics[width=15.5cm]{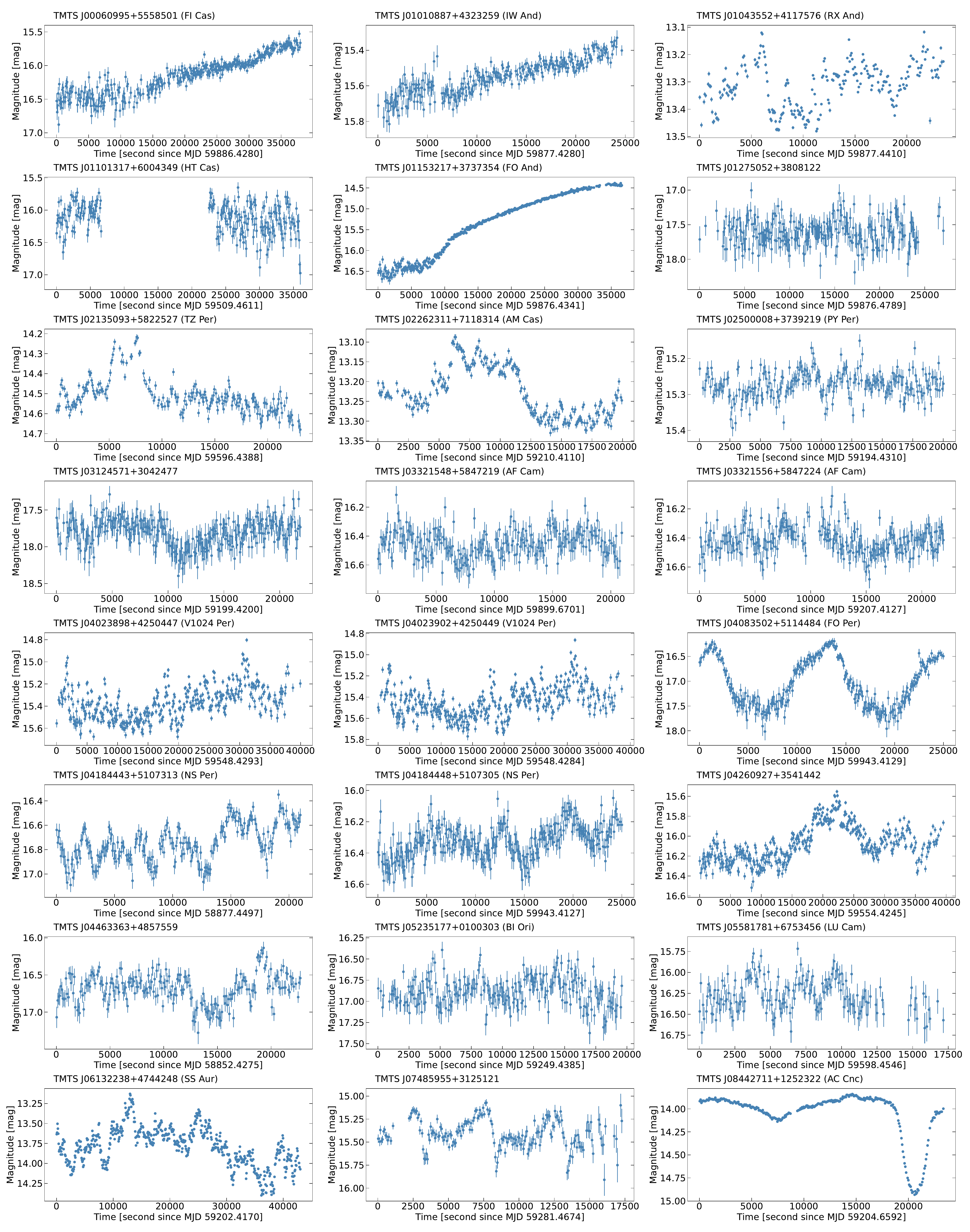}
 \end{adjustwidth}
	
\caption{\textit{Cont}.}
\end{figure}
\begin{figure}[H]\ContinuedFloat
 \begin{adjustwidth}{-\extralength}{0cm}
    \centering

	\includegraphics[width=15.5cm]{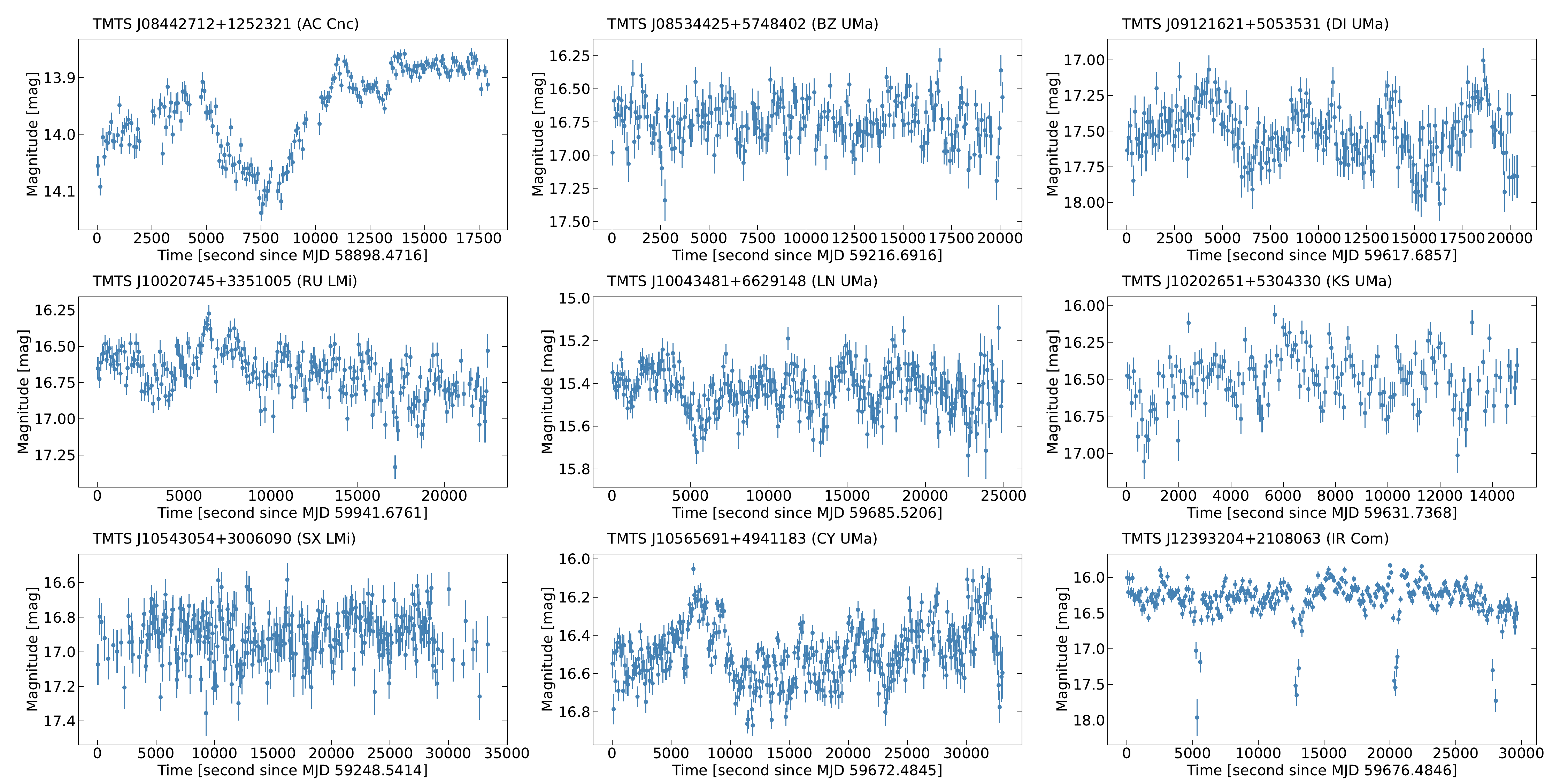}
	
 \end{adjustwidth}
    \caption{Light %MDPI: Please use commas to separate thousands for numbers with five or more digits (not four digits) in the picture, e.g., "10000" should be "10,000".
 curves of DNe presented in this work. The~start time of each observation can be found in Table~\ref{tab1}.  }
    \label{fig:lc}
\end{figure}
\unskip

%\begin{figure}[H]
%\ContinuedFloat
%	% To include a figure from a file named example.*
%	% Allowable file formats are eps or ps if compiling using latex
%	% or pdf, png, jpg if compiling using pdflatex
% \begin{adjustwidth}{-\extralength}{0cm}
%    \centering
% \end{adjustwidth}
%    \caption{(Continued).}
%    \label{fig:lc2}
%\end{figure}
%\unskip

\begin{figure}[H]
 \begin{adjustwidth}{-\extralength}{0cm}
    \centering
	\includegraphics[width=15.5cm]{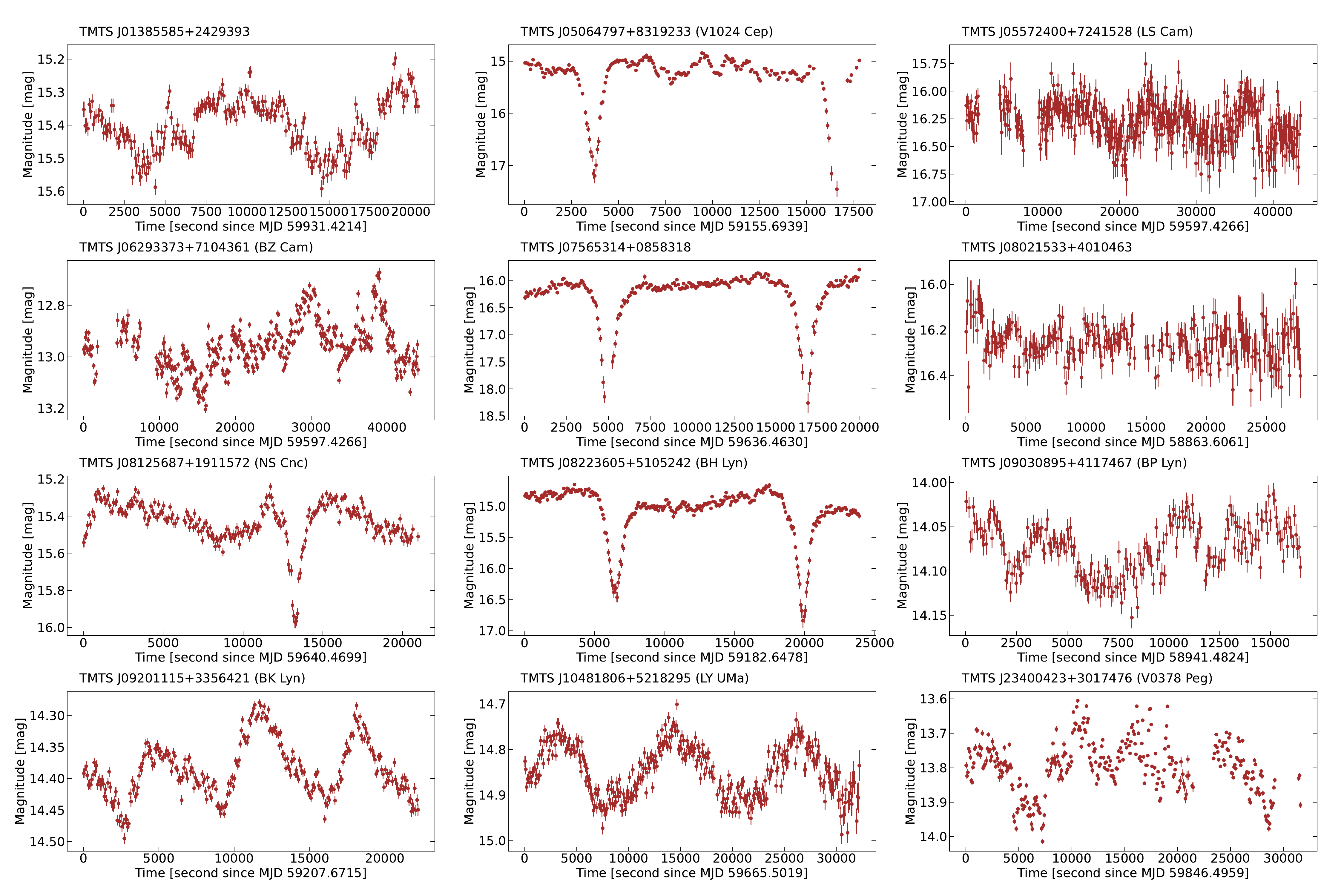}
 \end{adjustwidth}
    \caption{Light %MDPI: Please use commas to separate thousands for numbers with five or more digits (not four digits) in the picture, e.g., "10000" should be "10,000".
 curves of NLs presented in this work. The~start time of each observation can be found in Table~\ref{tab1}. }
    \label{fig:lc3}
\end{figure}
\unskip

\begin{figure}[H]
 \begin{adjustwidth}{-\extralength}{0cm}
    \centering
	\includegraphics[width=15.5cm]{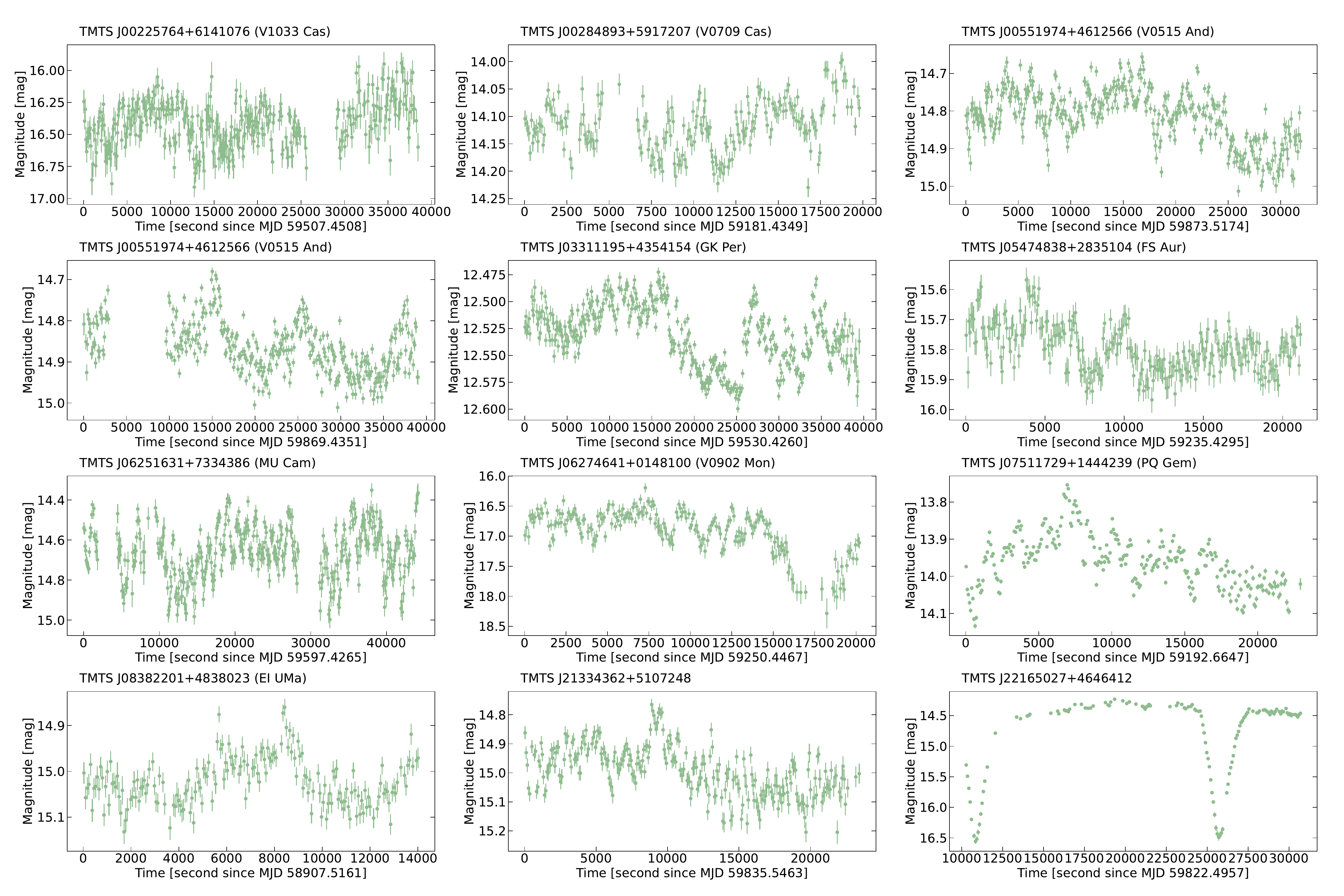}
 \end{adjustwidth}
    \caption{Light %MDPI: Please use commas to separate thousands for numbers with five or more digits (not four digits) in the picture, e.g., "10000" should be "10,000".
 curves of IPs presented in this work. The~start time of each observation can be found in Table~\ref{tab1}. }
    \label{fig:lc4}
\end{figure}
\unskip

\begin{figure}[H]
 \begin{adjustwidth}{-\extralength}{0cm}
    \centering
	\includegraphics[width=15.5cm]{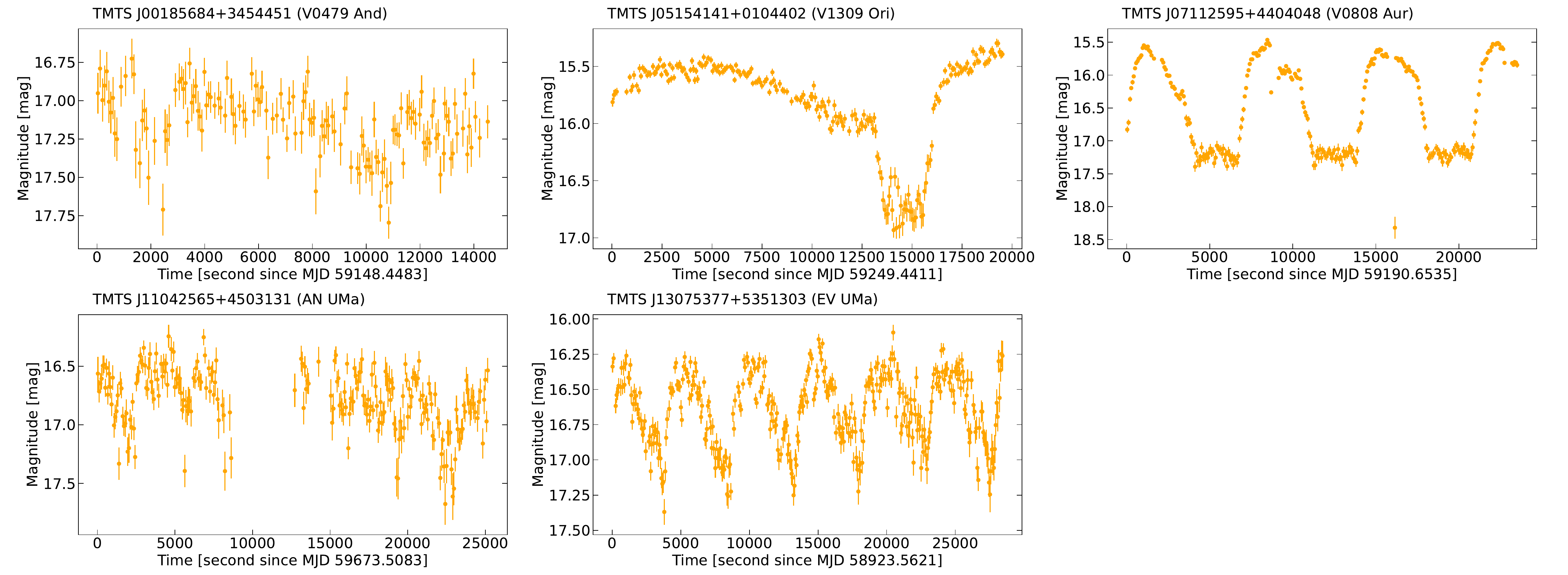}
 \end{adjustwidth}
    \caption{Light %MDPI: Please use commas to separate thousands for numbers with five or more digits (not four digits) in the picture, e.g., "10000" should be "10,000".
 curves of AMs presented in this work. The~start time of each observation can be found in Table~\ref{tab1}. }
    \label{fig:lc5}
\end{figure}
\unskip

\begin{figure}[H]
 \begin{adjustwidth}{-\extralength}{0cm}
    \centering
	\includegraphics[width=15.5cm]{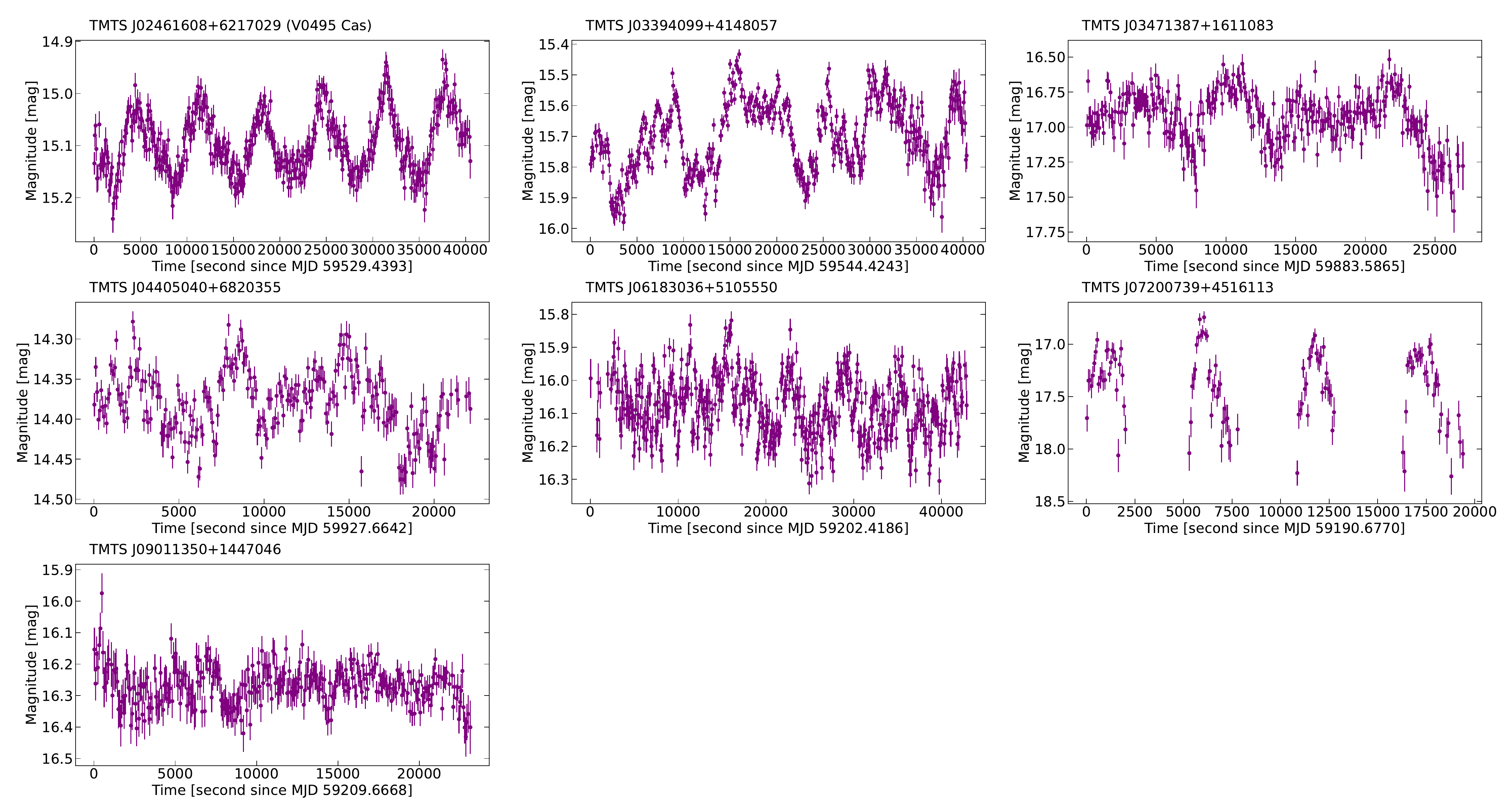}
 \end{adjustwidth}
    \caption{Light %MDPI: Please use commas to separate thousands for numbers with five or more digits (not four digits) in the picture, e.g., "10000" should be "10,000".
 curves of CV candidates presented in this work. The~start time of each observation can be found in Table~\ref{tab1}. }
    \label{fig:lc6}
\end{figure}
\unskip

\begin{figure}[H]
 \begin{adjustwidth}{-\extralength}{0cm}
    \centering
	\includegraphics[width=15.5cm]{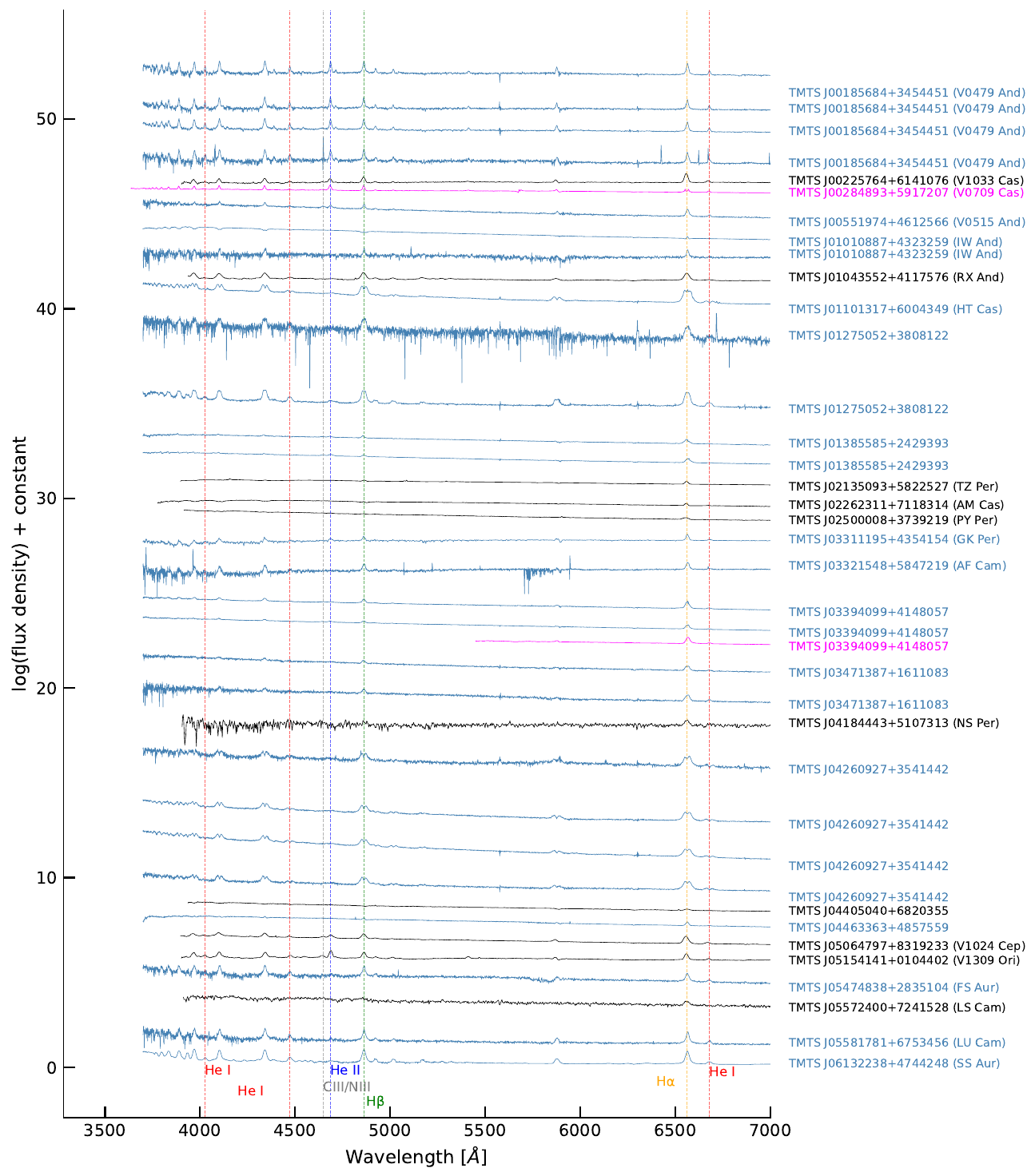}
 \end{adjustwidth}
	
\caption{\textit{Cont}.}
\end{figure}
\begin{figure}[H]\ContinuedFloat
 \begin{adjustwidth}{-\extralength}{0cm}
    \centering
	\includegraphics[width=15.5cm]{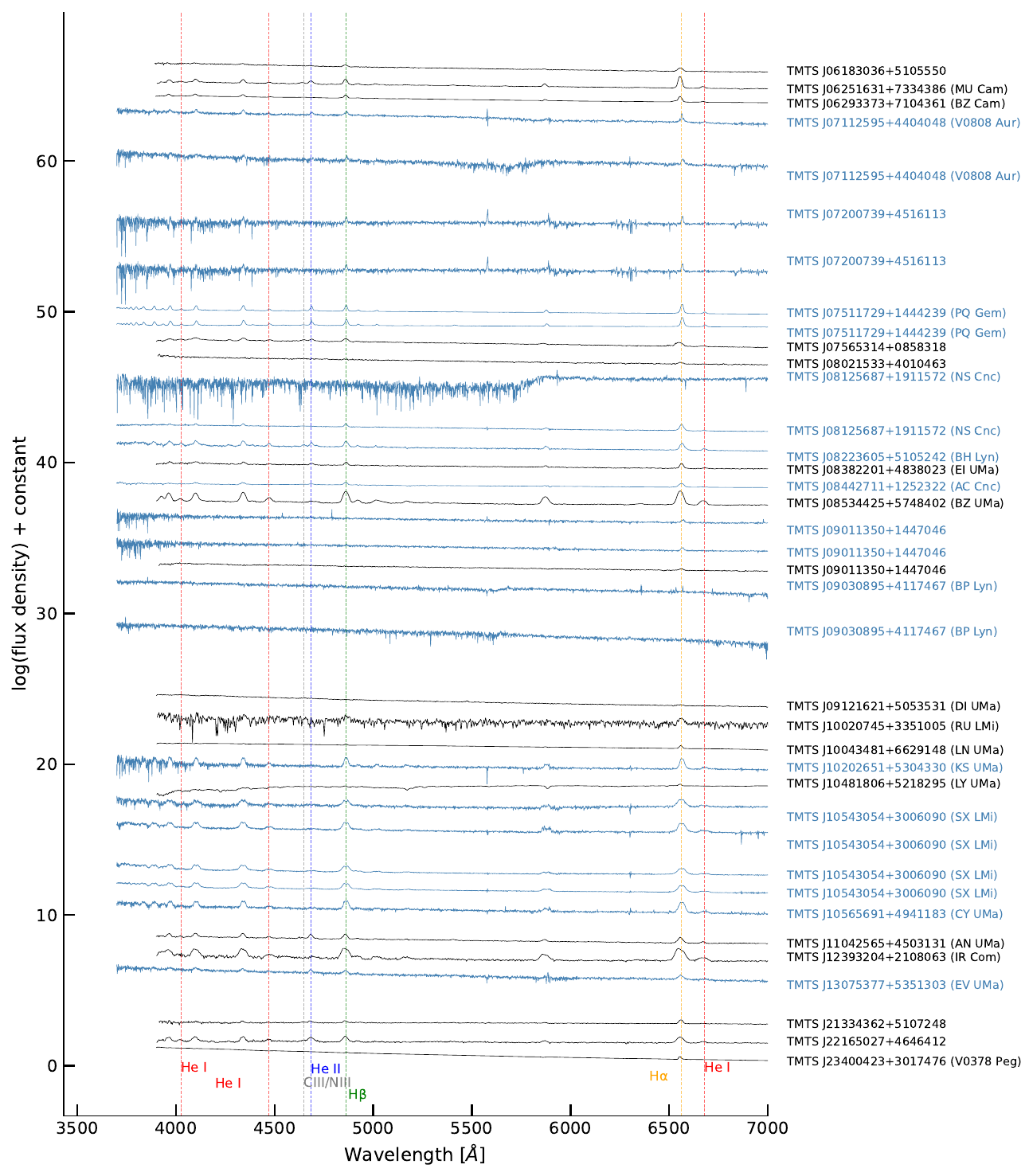}
 \end{adjustwidth}
    \caption{{Spectra} 
 of 53 cataclysmic variables mentioned in Section~\ref{spec}. The~spectra taken with the Xinglong 2.16~m telescope, LAMOST, and~Lick 3~m Shane telescope are shown in black, blue, and~magenta colors, respectively. Some spectral characteristics are indicated with vertical dashed lines in different colors. The~corresponding TMTS names are labeled on the right of each~spectrum.}
    \label{fig:spec}
\end{figure}

%\begin{figure}[H]
%\ContinuedFloat
%	% To include a figure from a file named example.*
%	% Allowable file formats are eps or ps if compiling using latex
%	% or pdf, png, jpg if compiling using pdflatex
% \begin{adjustwidth}{-\extralength}{0cm}
%    \centering
% \end{adjustwidth}
%    \caption{(Continued).}
%    \label{fig:spec2}
%\end{figure}

\begin{adjustwidth}{-\extralength}{0cm}

\reftitle{References}
%\bibliography{your_external_BibTeX_file}

\PublishersNote{}
\end{adjustwidth}
\end{document}